\documentclass[aps,pre,reprint,twocolumn,superscriptaddress,floatfix]{revtex4-2}

\usepackage{amsmath}
\usepackage{amsfonts}
\usepackage{amssymb}
\usepackage{mathtools}
\usepackage{empheq}
\usepackage{multirow}
\usepackage{indentfirst}
\usepackage{color}
\usepackage{float}

\usepackage{graphicx}
\usepackage[dvipsnames]{xcolor}
\usepackage{subcaption}

\usepackage[varg]{txfonts}
\usepackage{newtxtext}

\usepackage{bm}

\usepackage{physics}
\let\exp\exponential

\allowdisplaybreaks

\definecolor{pdarkblue}{rgb}{0.1797, 0.1875, 0.5703}
\usepackage{hyperref}
\hypersetup{
    colorlinks=true,
    citecolor=pdarkblue,
    linkcolor=pdarkblue,
    urlcolor=pdarkblue,
}
\usepackage[all]{hypcap}

\AtBeginDocument{
\fontsize{10.5pt}{12.36pt}\selectfont
\fontdimen2\font = 2.9pt 
}
\usepackage{titlesec}
\titleformat*{\section}{\centering\fontsize{10.5pt}{\baselineskip}\selectfont\bfseries}
\titleformat*{\subsection}{\centering\fontsize{10.5pt}{\baselineskip}\selectfont\bfseries}
\titleformat*{\subsubsection}{\centering\fontsize{10.5pt}{\baselineskip}\selectfont\itshape}
\titlespacing*{\section}{\linewidth}{2.5em}{0.4em}
\titlespacing*{\subsection}{\linewidth}{1.5em}{0.4em}
\titlespacing*{\subsubsection}{\linewidth}{1.5em}{0.4em}

\allowdisplaybreaks

\begin{document}

\title{Characteristic oscillations in frequency-resolved heat dissipation of linear time-delayed Langevin systems:  Approach from the violation of the fluctuation response relation}
\date{\today}

\author{Xin Wang}
\email{wangxin579@outlook.com}
\affiliation{Department of Physics, Graduate School of Science, The University of Tokyo, 5-1-5 Kashiwanoha, Kashiwa, Chiba 277-8574, Japan}

\author{Ruicheng Bao}
\affiliation{Department of Physics, Graduate School of Science, The University of Tokyo, 7-3-1 Hongo, Bunkyo-ku, Tokyo 113-0033, Japan}

\author{Naruo Ohga}
\affiliation{Department of Physics, Graduate School of Science, The University of Tokyo, 7-3-1 Hongo, Bunkyo-ku, Tokyo 113-0033, Japan}

\begin{abstract}
    Time-delayed effects are widely present in nature, often accompanied by distinctive nonequilibrium features, such as net heat flow from a single thermal bath to the system. To elucidate detailed structures of the dissipation, we study the frequency decomposition of the heat dissipation in linear time-delayed Langevin systems. We decompose the heat dissipation into frequency spectrum using the Harada-Sasa equality, which relates the heat dissipation to the violation of the fluctuation-response relation (FRR). We find a characteristic oscillatory behavior in the spectrum, and the oscillation asymptotically decays with an envelope inversely proportional to the frequency in the high-frequency region. Furthermore, the oscillation over the low-frequency region reflects the magnitude and sign of the heat dissipation.  We confirm the generality of the results by extending our analysis to systems with multiple delay times. Since the violation of FRR is experimentally accessible, our results suggest an experimental direction for detecting and analyzing detailed characteristics of dissipation in time-delayed systems. 
\end{abstract}

\maketitle

\section{Introduction}

Time delay is a paradigmatic non-Markovian mechanism that introduces memory effects into dynamical processes, a feature that is pervasive in both natural and engineered systems~\cite{FuncDiff,Optoelectronic,Feedback1,Feedback2,Feedback3,mechanical1,mechanical2,neural1,neural2,chemical,ecosystems,ecology2,biology1,biology2,biology3,biology4,biology5,biophys1,resonance1,resonance2,optical1,experiment1,experiment2,experiment3}.
These delays often arise from finite information-transmission times and finite response latencies.
For instance, in gene regulatory networks, delays arise from finite response speeds in processes such as transcription and translation, leading to stochastic oscillations in protein concentrations~\cite{biology4}. In neurodynamics, time delays in signaling between different nodes are critical for the emergence of slow oscillations and anticorrelated spatiotemporal patterns of cortical networks~\cite{biology5}. In mechanical and optoelectronic systems, properly designed control with time delay can improve the performance and enable applications such as stabilization by Pyragas control~\cite{Feedback1,Feedback2,Feedback3} and optical chaos communication~\cite{Optoelectronic,ChaosCom1,ChaosCom2}. Recent studies also utilize time-delayed feedback control of colloidal particles to simulate and investigate the collective behaviors of active matter~\cite{active1,active2,active3}.

In real-world scenarios, systems are often subject to noise due to random environmental perturbations or intrinsic fluctuations. From an energetic perspective, stochastic thermodynamics provides a powerful tool to quantify thermodynamic quantities like work and entropy production at the level of individual trajectories, and bridges these quantities to fluctuating statistics of dynamics~
\cite{sekimoto,FTreview,peliti2021}.
Key results include fluctuation theorems~\cite{Jarzynski1,Jarzynski2,Crooks1,Crooks2} that generalize the second law of thermodynamics and thermodynamic uncertainty relations~\cite{TUR,TURproof,TURreview} that bound the precision of currents with thermodynamic quantities. While these results were initially established for Markovian systems, the framework of stochastic thermodynamics has been successfully extended to various non-Markovian scenarios. These include dynamics with memory arising from coarse-grained degrees of freedom~\cite{coarse1,coarse2,coarse3,coarse4,coarse5,coarse6,coarse7,coarse8,coarse9}, as well as systems with time delay~\cite{munakata1, sarah1,experiment1,experiment2,sarah4,rosinberg1,sarah2,munakata2,rosinberg2,hasegawa1,Holubec,rosinberg3}. 
 For time-delayed systems, one of the remarkable behaviors is negative apparent heat dissipation, where the heat on average flows into the system from a single thermal bath~\cite{munakata1, sarah1,experiment1,experiment2,sarah4}. This can be understood through information-thermodynamic principles in the context of feedback control, where the apparent energy gain is offset by hidden costs associated with measurement and feedback processes~\cite{rosinberg1,sarah2}. Fluctuation theorems and thermodynamic uncertainty relations have also been generalized to time-delayed Langevin systems~\cite{munakata2,rosinberg2,hasegawa1,Holubec,rosinberg3}.

Despite these advances in understanding the thermodynamics of time-delayed systems,  most studies have primarily focused on building formal theoretical frameworks or deriving general principles.
A concrete and detailed analysis of thermodynamic features of time-delayed systems, beyond observing a single quantity such as negative heat dissipation, is crucial for understanding the time-delayed effects.
Moreover, discovering these thermodynamic features is also experimentally important as it can help us identify the time-delayed effects and effectively measure their thermodynamic behaviors.

In this study, we focus on linear time-delayed overdamped Langevin systems and analyze their characteristic properties of heat dissipation. Specifically, we use the Harada-Sasa equality~\cite{harada-sasa1} to decompose the heat dissipation rate into dynamical quantities in the frequency domain, which can be readily calculated from trajectory information, and study the detailed features of this frequency decomposition. We show that the spectrum resolved from the heat dissipation rate manifests a typical oscillation over the entire frequency domain. The shape of the spectrum at low frequencies largely determines the sign and the magnitude of the heat dissipation rate. In the high-frequency limit, the spectrum exhibits a sinusoidal oscillation whose frequency corresponds to the delay time $\tau$, with a decaying envelope of scaling $1/\omega$.  These properties highlight characteristics of linear time-delayed systems, which suggests a direction to identify and analyze time-delayed effects in experimental systems.

This paper is organized as follows. In Sec.~\ref{section:2}, we introduce a general class of Langevin systems with linear time-delayed forces and review the Harada-Sasa equality for general time-delayed Langevin systems.  In Sec.~\ref{section:3}, we analytically solve the dynamics of linear Langevin systems with a single delay term and calculate the heat dissipation rate and 
its frequency-resolved spectrum. We discuss the behaviors of the spectrum in detail.
In Sec.~\ref{section:4}, we extend our results to the systems with multiple delay times. Section~\ref{section:conclusion} concludes the paper.

\section{Setup and the Harada-Sasa equality}\label{section:2}
We start with a general class of time-delayed systems described by a stochastic delay differential equation (SDDE):
\begin{equation}\label{gen_eq}
  \left\{
    \begin{alignedat}{2}
      &m\ddot{x}(t) = -\gamma \dot{x}(t) + F(x(t),x(t-\tau_1),\cdots,x(t-\tau_n ))    \\
      &\quad \quad\quad   + \xi(t) + \epsilon f^p(t),
      & (t>t_{\text{init}}), \\ 
      &x(t) = \phi(t),  &\hspace{-5.5em} (t\in [t_{\text{init}}-\tau_{\mathrm{max}} ,t_{\text{init}}]),  \\ 
      &\dot{x}(t_{\text{init}}) = v_0.  &\hspace{-2.5em}   \\ 
    \end{alignedat}
  \right.
\end{equation}
Here, $m$ and $\gamma$ denote the mass and friction coefficient, respectively. The term $F(x(t),x(t-\tau_1),\cdots,x(t-\tau_n ))$ includes both time-local and time-delayed forces in the dynamics. We assume delay times $\tau_1,\cdots,\tau_n$ are constants and $\tau_{\mathrm{max}}=\max\{\tau_1,\cdots,\tau_n\}$. The term $\xi(t)$ is the white Gaussian noise with mean $\left\langle \xi(t)\right\rangle =0$ and correlation $\left\langle \xi(t)\xi(t')\right\rangle = 2\gamma T \delta (t-t')$, where $T$ is the temperature of the surrounding heat bath. Throughout this study, we assume that the delay time $\tau$ is sufficiently longer than the physical timescale of the environmental noise correlation, thus the noise can be safely modeled as uncorrelated in time.  We also add a perturbation force $\epsilon f^p(t)$ with a small strength $\epsilon$ to define the response function below. The dynamics starts at $t=t_{\text{init}}$ with a specified historical trajectory $\phi(t)$ over $t\in [t_{\text{init}}-\tau_{\mathrm{max}} ,t_{\text{init}}]$ and an initial velocity $v_0$.  
The SDDE dynamics apply to various fields including ecology~\cite{ecology2}, biophysics~\cite{biophys1}, and engineering~\cite{optical1,mechanical1,mechanical2}. In experiments, such time-delayed dynamics can also be implemented using an optically levitated particle under time-delayed feedback control in underdamped~\cite{experiment1,experiment2} and overdamped~\cite{experiment3,active3} cases.

The dynamics may converge to a stationary process or diverge indefinitely depending on the functions $F(x(t),x(t-\tau_1),\cdots,x(t-\tau_n ))$ and initial conditions. In the following, we assume the convergence and take the limit $t_\mathrm{init}\to-\infty$ to focus on the steady state. We use $\left\langle \cdots \right\rangle_\epsilon$ to express the steady-state average over all possible realizations of trajectories with the perturbation strength $\epsilon$. 

The stochastic heat dissipation rate $J(t)$ is defined through
\begin{equation}\label{heat_def}
  J(t)dt \equiv [\gamma \dot{x}(t)-\xi(t)]\circ d x(t),
\end{equation}
where $\circ$ represents the Stratonovich product~\cite{sekimoto}. Here, $-\gamma \dot{x}(t)+\xi(t)$ represents the friction force and the random thermal force applied to the system by the environment. Therefore, $J(t)dt$ represents the work done by the system’s counterforce on the environment over the displacement  $d x$ within time $d t$ at the level of a single trajectory. This energy transfer is interpreted as heat dissipation. When this heat dissipation rate is positive ($J(t) > 0$), energy flows from the system to the environment. The Stratonovich product in Eq.~\eqref{heat_def} ensures the validity of the first law of thermodynamics (see Ref.~\cite{sekimoto} for details).   
For Markovian systems that satisfy Einstein’s relation, the average heat dissipation rate is always non-negative, consistent with the second law of thermodynamics. However, time-delayed systems may have a negative heat dissipation rate on average. This is compensated by extra heat dissipation in the external degrees of freedom that exert the time-delayed force to the system, which
are not modeled explicitly in Eq.~\eqref{gen_eq}. For example, the extra heat dissipation may arise due to the thermodynamic cost associated with the recorded information in 
the feedback-control agency~\cite{experiment1, rosinberg1,Landauer,sagawa}.

This heat dissipation rate is related to two dynamical quantities of the system, as originally reported in Ref.~\cite{harada-sasa1}. The first quantity is the response function $R_v(t)$ in the steady state, defined as
\begin{equation}\label{response_fun}
  R_v(t-t')\equiv 
  \dfrac{\delta \left\langle \dot{x}(t)\right\rangle_\epsilon }{\delta [\epsilon f^p(t')]}\bigg|_{\epsilon=0}
\end{equation}
which yields
\begin{equation}\label{response}
  \left\langle \dot{x}(t)\right\rangle _{\epsilon}= v_s+ \epsilon \int_{-\infty}^{t}R_v(t-s)f^p(s)ds+O(\epsilon ^2),
\end{equation}
where $\delta /\delta [\epsilon f^p(t)]$ represents the functional derivative and $v_s\equiv \langle \dot{x}(t) \rangle _0$ is the steady-state velocity independent of $t$.
Notice that $R_v(t)=0$ when $t<0$ because of causality.  The other quantity is the time-correlation function of velocity in the absence of perturbation in the steady state,
\begin{equation}\label{corr_fun}
  C_v(t)\equiv \left\langle \left[ \dot{x}(t)-v_s\right] \left[ \dot{x}(0)-v_s \right]\right\rangle_0,
\end{equation}
which satisfies $C_v(-t)=C_v(t)$.

The Harada-Sasa equality relates these two quantities to the heat dissipation rate in the steady state by the following identity~\cite{harada-sasa1}:
\begin{equation}\label{hs}
\left\langle J \right\rangle_0 = \gamma \left\{ v_s^2 + \int_{-\infty}^{\infty}\left[\tilde{C}_v(\omega)-
2T\tilde{R}'_v(\omega)\right] \frac{d\omega}{2\pi}\right\}.   
\end{equation}
Here and throughout this paper, we define the Fourier transform of a function $f(t)$ as $\tilde{f}(\omega)\equiv \int_{-\infty}^{\infty} f(t)\exp(i\omega t)dt$ and $\tilde{R}'_v(\omega)$ denotes the real part of $\tilde R_v(\omega)$.  Notice that $\tilde{C}_v(\omega)$ represents the power spectral density of the velocity and $\epsilon \tilde{R}_v(\omega)$ corresponds to the velocity response to the specific perturbation force $\epsilon f^p(t) =\epsilon e^{i\omega t}$. The fluctuation-response relation (FRR) $\tilde{C}_v(\omega)=2T \tilde{R}'_v(\omega)$ always holds for all frequencies in equilibrium systems~\cite{kubo,harada-sasa2}, and thus the response and the power spectral density are always balanced. In contrast, FRR is generally violated in nonequilibrium systems. The Harada-Sasa equality shows that the FRR violation corresponds to a positive (negative) contribution to the heat dissipation rate if the power spectral density is stronger (weaker) than the response.
 The Harada-Sasa equality was first proved for overdamped Langevin systems~\cite{harada-sasa1}, and then generalized to underdamped Langevin systems~\cite{harada-sasa2,ohkuma}, systems with colored noise~\cite{harada-sasa2,deutsch}, density-field variables~\cite{harada3}, quantum systems~\cite{saito},  and active matter~\cite{active-matter1,active-matter2}. It has been used to experimentally measure heat dissipation in Brownian particles~\cite{HS-experimental1,HS-experimental3} and biological systems~\cite{HS-experimental2,HS-experimental4,HS-experimental5}.

The Harada-Sasa equality holds for general time-delayed systems in Eq.~\eqref{gen_eq}. The proof of Eq.~\eqref{hs} for time-delayed systems can be done in exactly the same way as the proof for Markovian Langevin systems in Ref.~\cite{harada-sasa2}. Notably, our setup is regarded as a special case of the setup in Ref.~\cite{ito}, and thus Eq.~\eqref{hs} is a special case of the Harada-Sasa equality proved in Ref.~\cite{ito}. For completeness, we provide a proof of Eq.~\eqref{hs} in Appendix \ref{appendix:Harada-Sasa}.

\section{Linear systems with a single delay time}\label{section:3}

In this section, we focus on a simple yet insightful class of time-delayed systems described by the following overdamped Langevin equation:
\begin{equation}\label{par_eq}
  \left\{
    \begin{alignedat}{2}
      &\gamma \dot{x}(t) = ax(t) + bx(t-\tau) + \epsilon f^p(t) + \xi(t) ,
      & (t>t_{\text{init}}), \\ 
      &x(t) = \phi(t),  &\hspace{-2em}(t\in [t_{\text{init}}-\tau ,t_{\text{init}}]),  \\ 
    \end{alignedat}
  \right.
\end{equation}
where $a$ is the strength of a linear force without time delay, $b$ is the strength of a linear time-delayed force, and $\tau$ denotes a single constant delay time ($\tau>0$). We assume that the other parameters share the same meaning as in Eq.~\eqref{gen_eq}. 
 The analytical solution to Eq.~\eqref{par_eq} has been provided by K{\"u}chler and Mensch~\cite{kuchler1992}, which reads 
\begin{align}\label{formal_sol}
  x(t) &= x_0(t-t_{\text{init}})\phi(t_{\text{init}})+
  \frac{b}{\gamma}\int_{t_{\text{init}}-\tau}^{t_{\text{init}}} x_0(t-s-\tau)\phi(s)ds \nonumber \\
  & \quad + \frac{1}{\gamma}\int_{t_{\text{init}}}^t x_0(t-s) [ \xi(s) + \epsilon f^p(s) ] ds,
\end{align} 
where $x_0$ is the fundamental solution, generally defined as the solution to the deterministic version of Eq.~\eqref{par_eq} 
satisfying the following conditions:
\begin{equation}
  \label{fundamental}
    \epsilon =0,\hspace{0.5em}
    T=0,\hspace{0.5em}
    t_\mathrm{init}=0,\hspace{0.5em}
    \phi(t)=0\,\,(t<0),\hspace{0.5em}
    \phi(0)=1.
\end{equation}
For the linear dynamics with single delay [Eq.~\eqref{par_eq}], $x_0(t)$ reads~\cite{kuchler1992}
\begin{equation}
  x_0(t) = \sum_{k=0}^{\lfloor t/\tau\rfloor}\frac{(b/\gamma)^k}{k!}(t-k\tau)^k e^{(a/\gamma)(t-k\tau)}.
\end{equation}
Notice that the first two terms of Eq.~\eqref{formal_sol} are history-dependent deterministic functions in general, while the last term corresponds to a history-independent function with a Gaussian distribution.

\begin{figure}[tb]
  \centering
  \begin{subfigure}[b]{0.23\textwidth}
    \includegraphics[width=\textwidth]{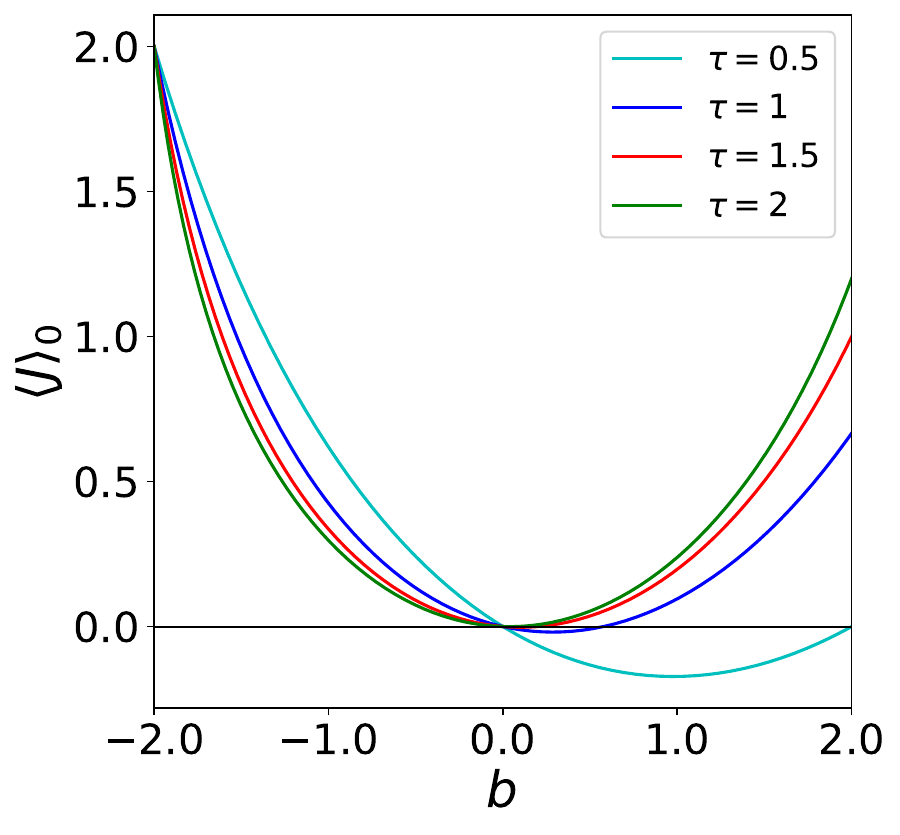}
    \caption{$a=-2$}
    \label{fig:image1}
  \end{subfigure}
  \hfill
  \begin{subfigure}[b]{0.23\textwidth}
    \includegraphics[width=\textwidth]{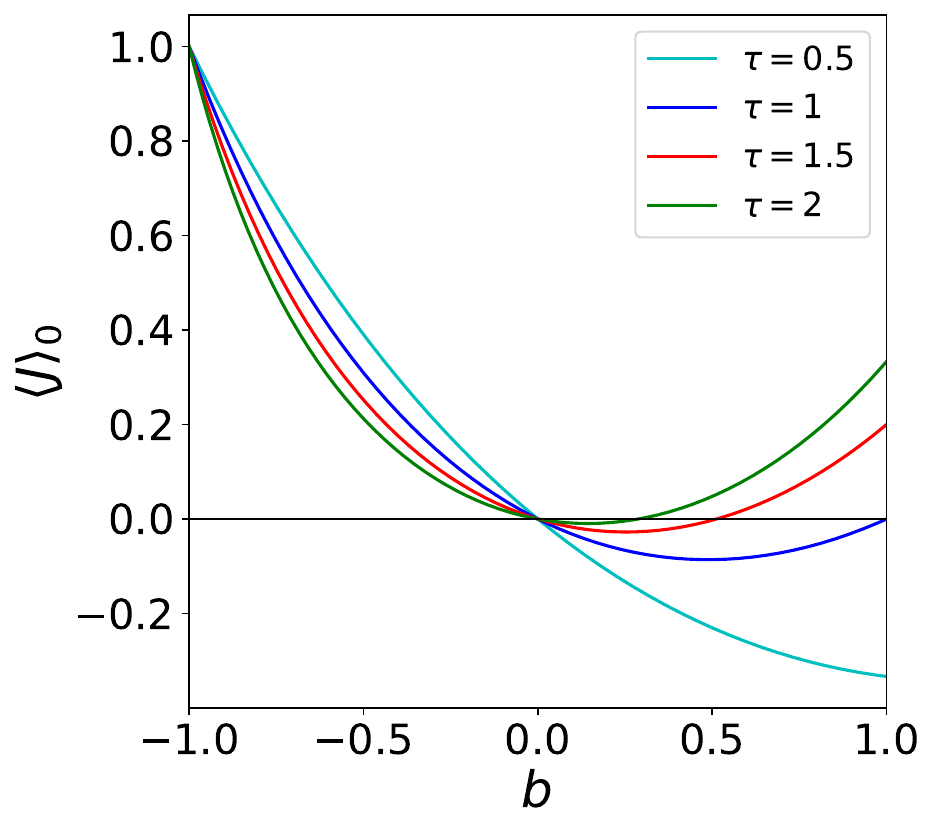}
    \caption{$a=-1$}
    \label{fig:image2}
  \end{subfigure}
  \\
  \begin{subfigure}[b]{0.23\textwidth}
    \includegraphics[width=\textwidth]{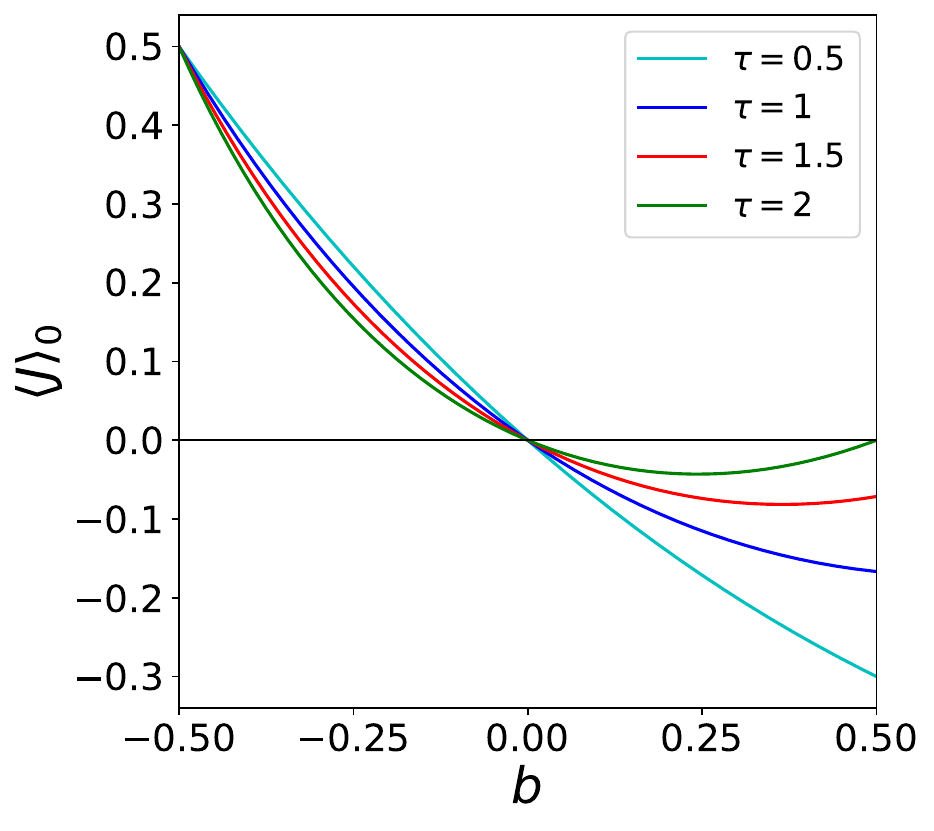}
    \caption{$a=-0.5$}
    \label{fig:image3}
  \end{subfigure}
  \hfill
  \begin{subfigure}[b]{0.23\textwidth}
    \includegraphics[width=\textwidth]{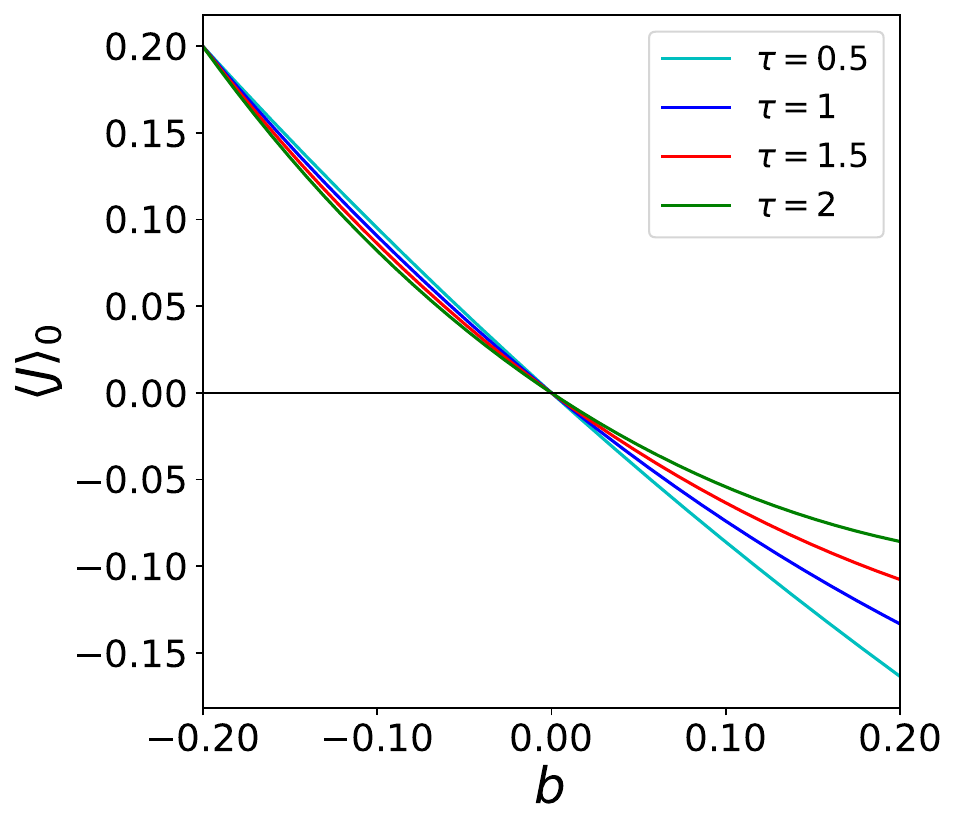}
    \caption{$a=-0.2$}
    \label{fig:image4}
  \end{subfigure}
  \caption{\raggedright The heat dissipation rate $\langle J\rangle_0$ for different values of $a$, $b$, and $\tau$ (here, $a<0$ and $|b|<|a|$). We fix $T=1$ and $\gamma=1$.}
  \label{fig:1}
\end{figure}

In the following, we assume that 
the solution converges to a stationary process in the long-time limit. The mathematical condition for the convergence is either
\begin{equation}\label{convergence_condition1}
   a+b<0 \ \ \ \text{and} \ \ \ a-b\leq0
\end{equation}
with an arbitrary delay time $\tau>0$, or 
\begin{equation}\label{convergence_condition2}
   a+b<0, \ \  \ \ a-b>0, \ \ \text{and} \ \  0<\tau<\dfrac{\gamma\arccos(-a/b)}{(b^2-a^2)^{1/2}}, 
\end{equation}
as shown in Ref.~\cite{kuchler1992}.  
In these cases, $\lim_{t\rightarrow \infty}x_0(t)=0$ and the stationary distribution in the long-time limit has zero mean position and velocity:
\begin{equation}
    \lim_{t\rightarrow \infty}\langle x(t)\rangle=0, \ \ v_s=\lim_{t\rightarrow\infty}\langle\dot x(t)\rangle=0.
\end{equation}

From the solution in Eq.~\eqref{formal_sol}, we can readily calculate the quantities of interest in the steady state (see Appendix \ref{appendix:CR_t} for the derivation). The linear response function and the time-correlation function in the steady state read
\begin{align}
    R_v(t) &= \frac{1}{\gamma^2} \left[ax_0(t)+bx_0(t-\tau)\right] + \frac{1}{\gamma} \delta(t),  \label{R} \\
    C_v(t) &= \frac{2T}{\gamma^3} \bigl\{ (a^2+b^2)K(t)+ab\left[K(t+\tau)+K(t-\tau)\right] \bigr\}
    \nonumber \\
    &\quad  + \frac{2T}{\gamma^2} [ ax_0(|t|)+bx_0(|t|-\tau)] + \frac{2T}{\gamma } \delta(t),  \label{C}
\end{align}
where we define $K(t)\equiv \int_0^\infty x_0(s)x_0(s+t)ds$, whose explicit formula is provided in Appendix \ref{appendix:CR_t}\@.  
We can also obtain the average heat dissipation rate for $\tau>0$:
\begin{equation}\label{heat}
\left\langle J \right\rangle_0=\frac{T}{\gamma} a+ \frac{2T}{\gamma^2}[ ( a^2 + b^2 )  K(0) + 2ab K(\tau) ].     
\end{equation}
Note that this expression of the heat dissipation rate has been derived in Refs.~\cite{munakata1,sarah1} from the time-delayed Fokker-Planck equation~\cite{frank2003}. In the vanishing delay limit ($\tau \to 0$), the expression of $\langle J\rangle_0$ in Eq.~\eqref{heat} approaches $-bT/\gamma$. However, when $\tau$ is exactly 0, Eq.~\eqref{heat} is no longer valid, and the heat dissipation rate strictly becomes zero since the heat dissipation is purely due to the time-delay force in the present setup.
Thus, $\langle J\rangle_0$ shows a discontinuity in the limit $\tau \to 0$ in the current overdamped model~\cite{munakata1,sarah1}. Such unphysical picture of the discontinuity originates from the failure of $\langle \xi(t)\xi(t-\tau)\rangle=0$ and $\langle x(t-\tau)\xi(t)\rangle=0$ in deriving Eq.~\eqref{heat} when the delay time $\tau$ is comparable to the timescale of the environmental noise correlation in physical scenarios (see Appendix~\ref{appendix:CR_t} and Chapter 9 of Ref.~\cite{sarah2} for related discussion).


The heat dissipation rate may become negative depending on the parameters $a$, $b$, and $\tau$. For example, we show the curves of heat dissipation with a restoring nondelayed linear force ($a<0$) in Fig.~\ref{fig:1}. Specifically, the heat dissipation rate can be negative for moderate delay times with a positive delayed force ($b>0$), while it always remains positive with a negative delayed force ($b<0$). When $a$ and $b$ are fixed, a shorter delay time results in lower (higher) heat dissipation rate when $b>0$ ($b<0$). The phenomena of negative heat dissipation have also been mentioned in several studies related to time-delayed feedback cooling~\cite{experiment1, experiment2,  rosinberg1, sarah1,sarah4}. In particular, Ref.~\cite{sarah4} reported similar curves of heat dissipation of a linear system with a delta-distributed memory kernel.

As the main result of the present paper, we analyze the heat dissipation rate in more detail by decomposing it into the frequency components using the Harada-Sasa equality. From Eqs.~\eqref{R} and \eqref{C}, we can analytically calculate the Fourier transforms $\tilde{C}_v(\omega)$, $\tilde{R}_v'(\omega)$ and the spectral decomposition of the heat dissipation  $\tilde{C}_v(\omega)-2T\tilde{R}'_v(\omega)$ (see Appendix \ref{appendix:CR_omega} for the derivation):
\begin{align}
  &\tilde{R}_v'(\omega)= -\frac {1}{\gamma} \frac{a^2+b^2 + 2ab\cos(\omega\tau) +b\gamma\omega \sin(\omega\tau) }{[a+b\cos(\omega\tau)]^2+[\gamma\omega+b\sin(\omega\tau)]^2} + \frac{1}{\gamma},  \\
  &\tilde{C}_v(\omega) = -\frac{2T}{\gamma} \frac{a^2 + b^2 + 2ab \cos(\omega\tau) + 2b\gamma\omega \sin(\omega \tau )} {[a+b\cos(\omega\tau)]^2+[\gamma\omega+b\sin(\omega\tau)]^2}  
    \!+\!\frac{2T}{\gamma},   \\
  &\tilde{C}_v(\omega)-2T\tilde{R}'_v(\omega) 
  \nonumber \\
  & \quad =  -2T \frac{ b\omega \sin(\omega \tau )} {[a+b\cos(\omega\tau)]^2+[\gamma\omega+b\sin(\omega\tau)]^2}.  \label{spectrum}
\end{align}

\begin{figure*}
  \centering
  \begin{subfigure}[b]{0.41\textwidth}
    \includegraphics[width=\textwidth]{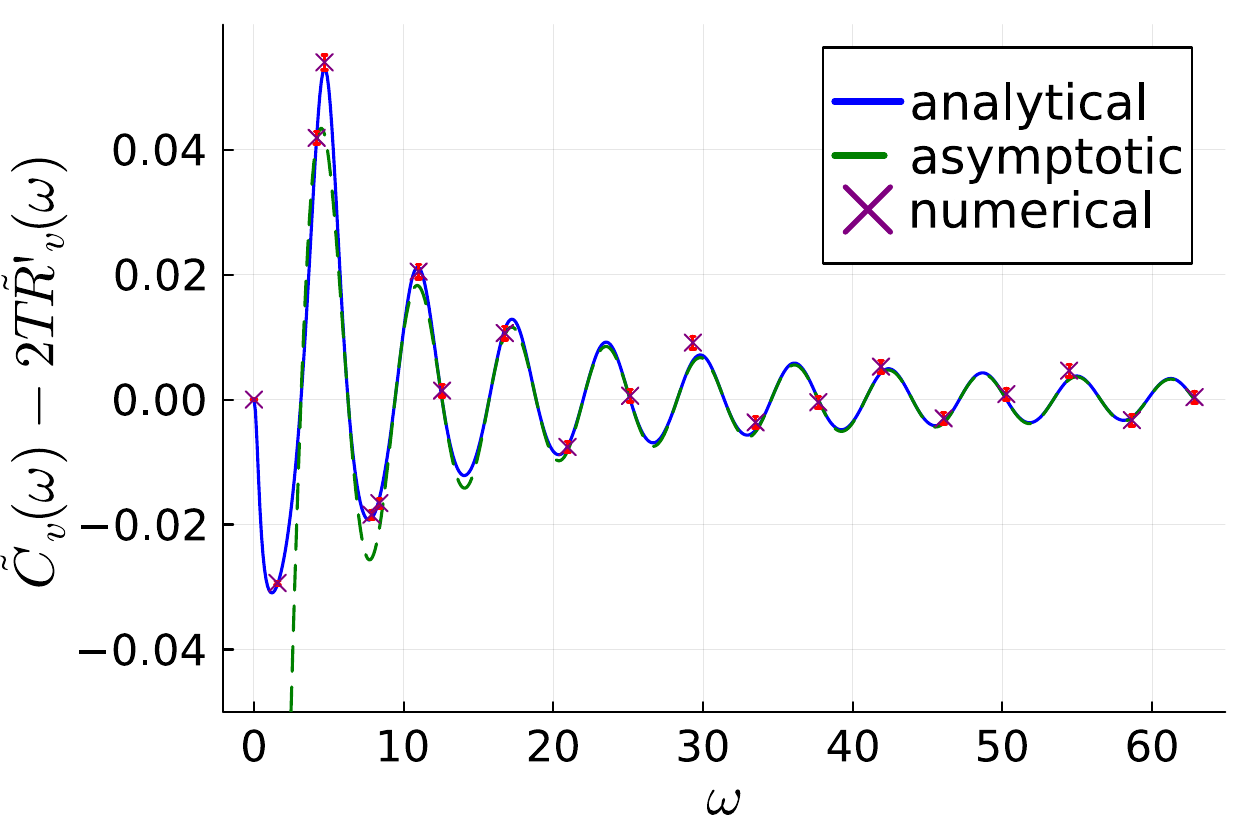}
    \caption{}
    \label{fig:2a}
  \end{subfigure}
  \hspace{5em}
 \begin{subfigure}[b]{0.41\textwidth}
    \includegraphics[width=\textwidth]{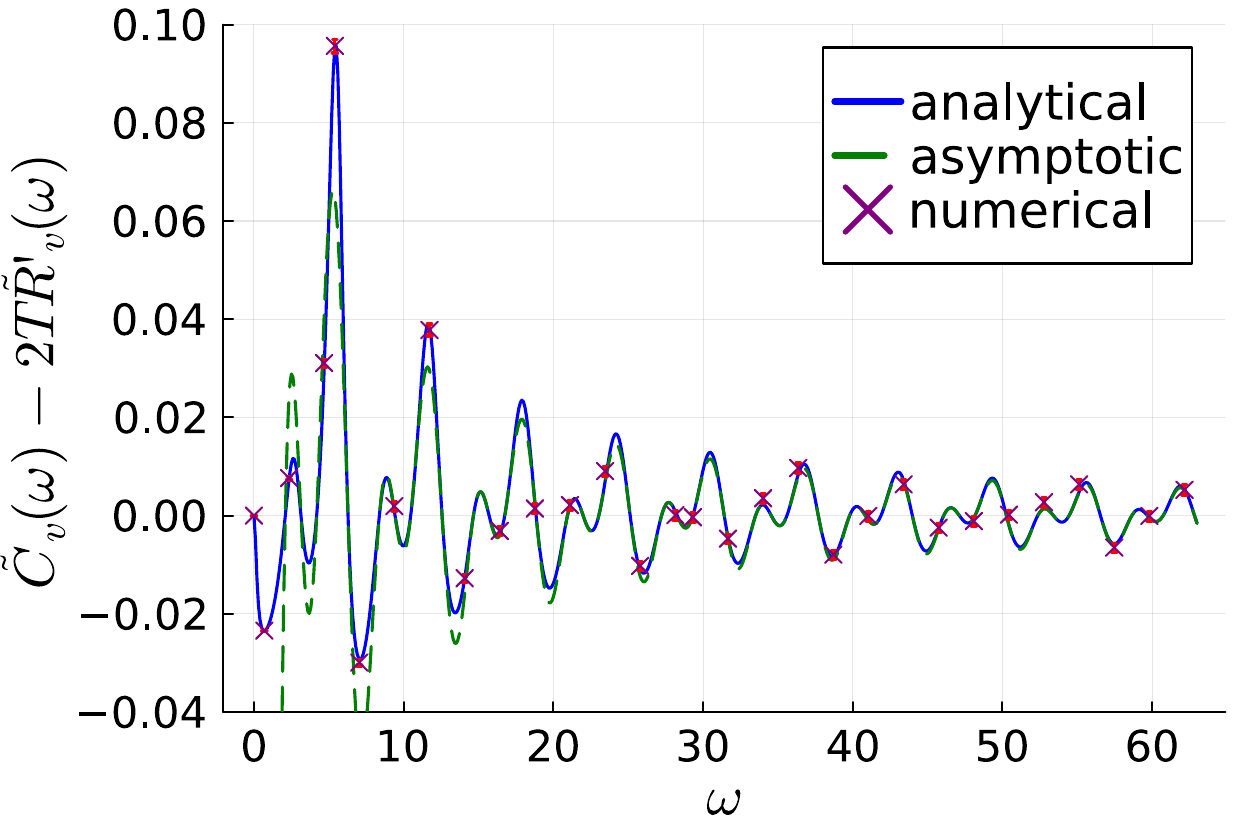}
    \caption{}
    \label{fig:2b}
  \end{subfigure}

   \par\vspace{0.3cm}
   
  \begin{subfigure}[b]{0.325\textwidth}
    \includegraphics[width=\textwidth]{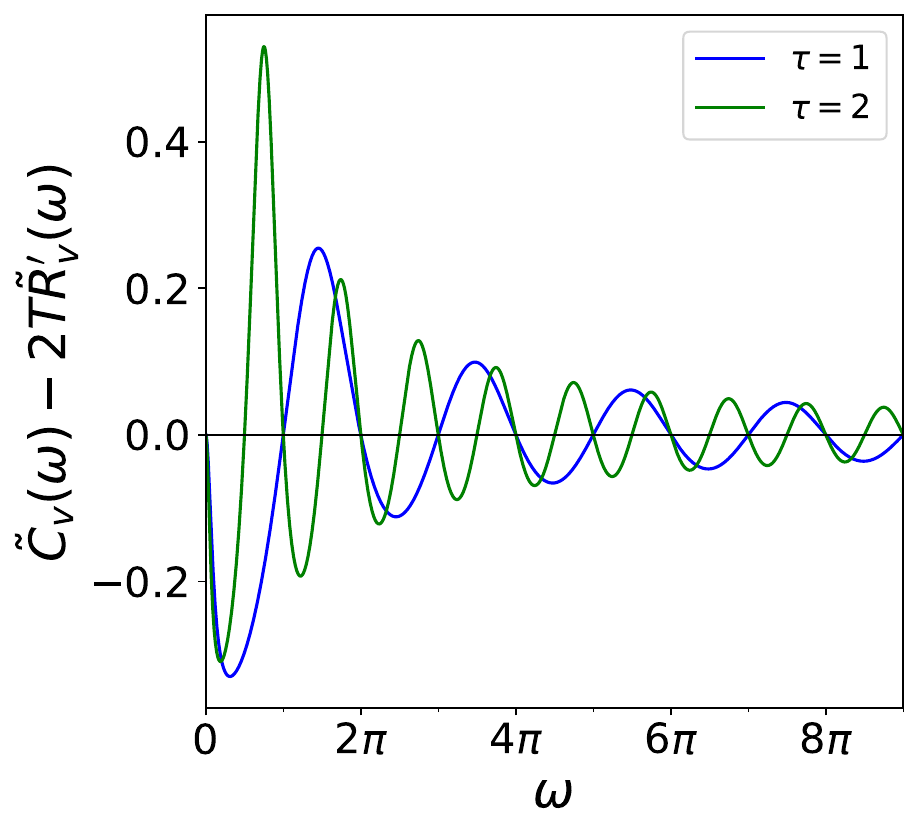}
    \caption{}
    \label{fig:2c}
  \end{subfigure}
 \hspace{5em}
  \begin{subfigure}[b]{0.325\textwidth}
    \includegraphics[width=\textwidth]{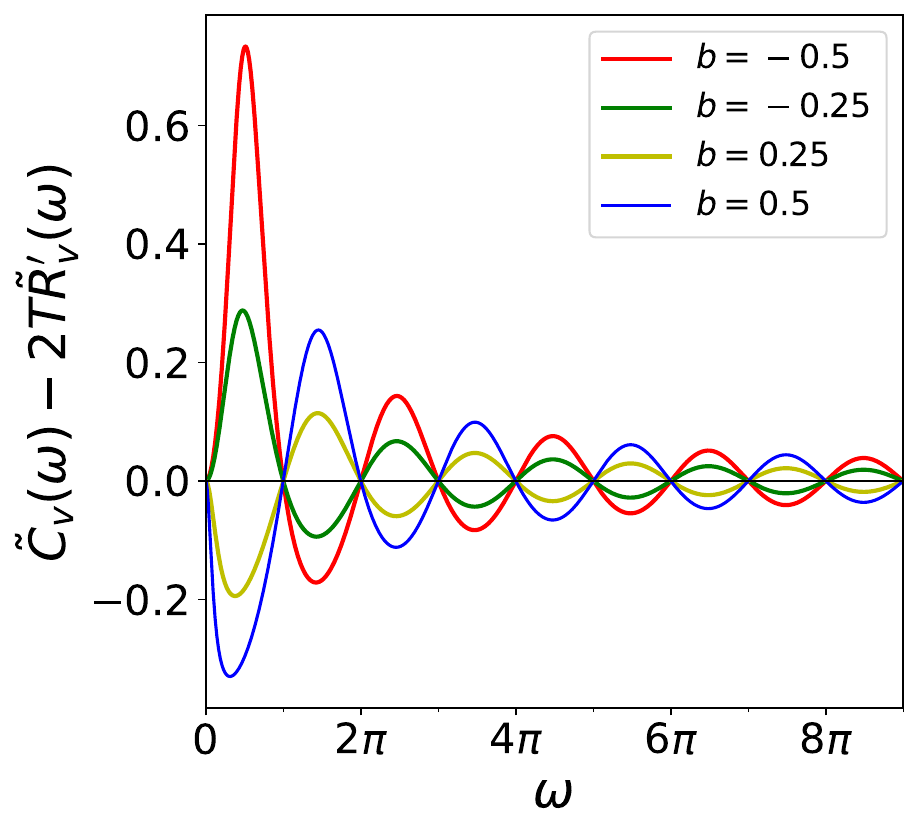}
    \caption{}
    \label{fig:2d}
  \end{subfigure}

 \par\vspace{0.3cm}


  \begin{subfigure}[b]{0.325\textwidth}
    \includegraphics[width=\textwidth]{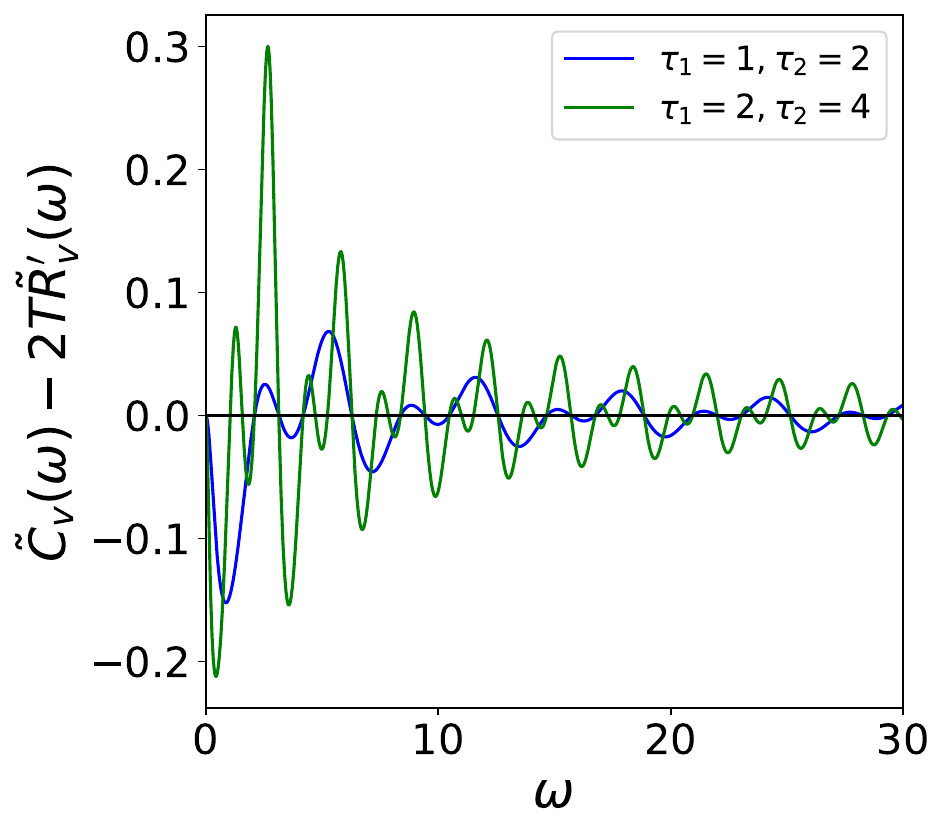}
    \caption{}
    \label{fig:2e}
  \end{subfigure}
  \hfill
  \begin{subfigure}[b]{0.325\textwidth}
    \includegraphics[width=\textwidth]{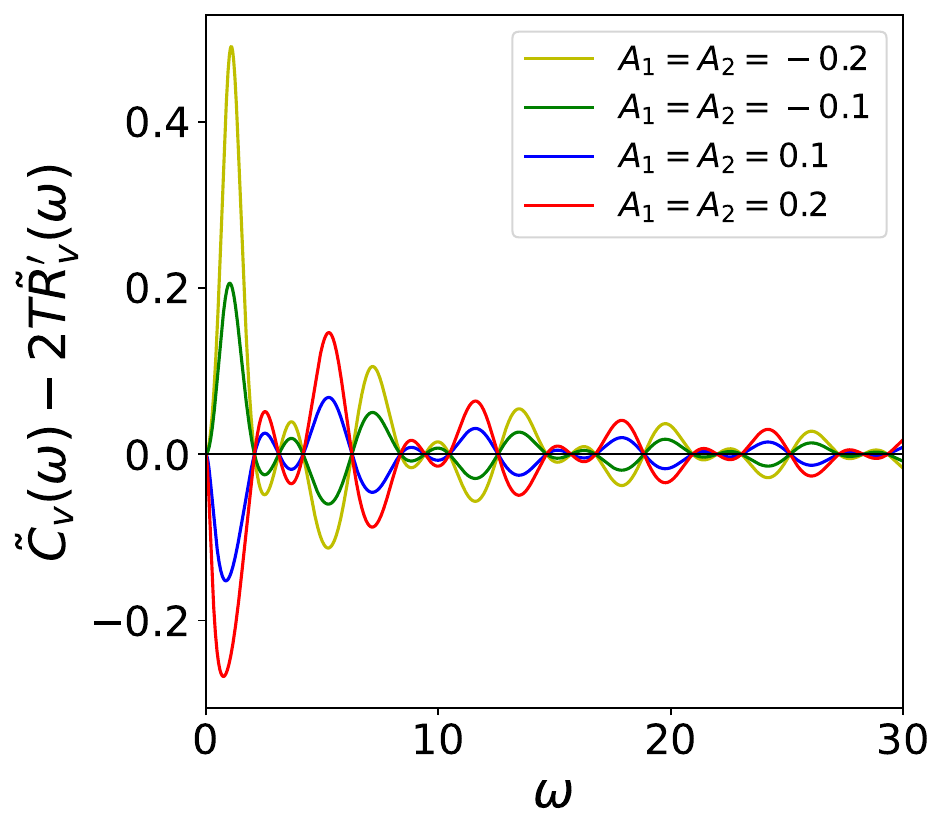}
    \caption{}
    \label{fig:2f}
  \end{subfigure}
  \hfill
  \begin{subfigure}[b]{0.325 \textwidth}
    \includegraphics[width=\textwidth]{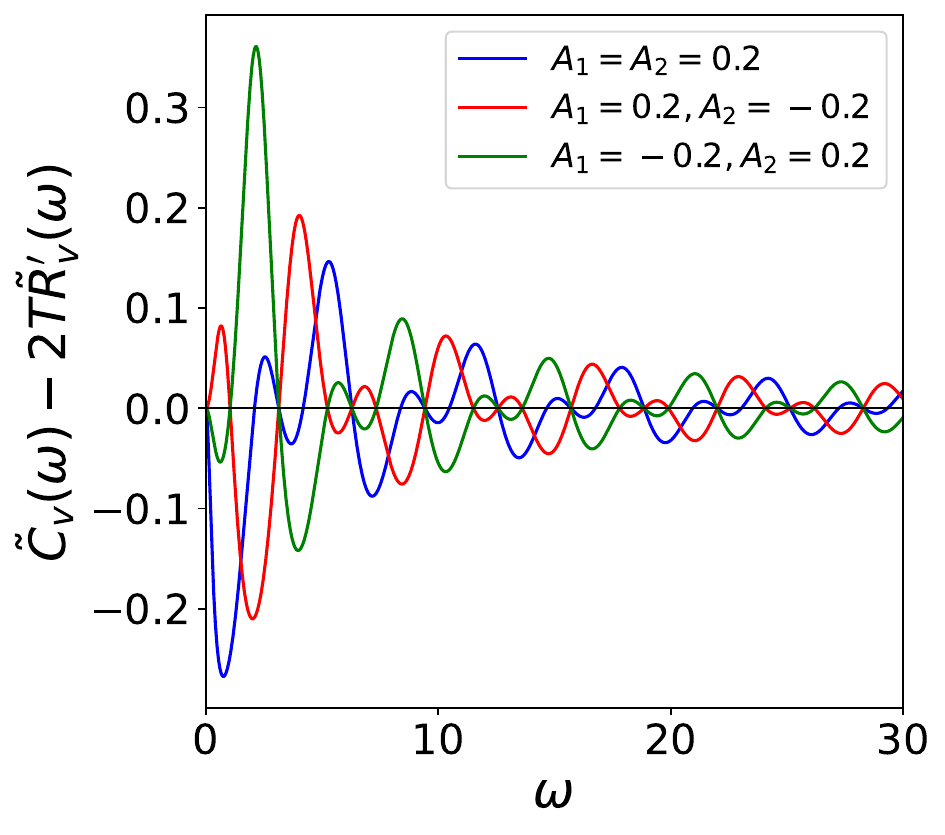}
    \caption{}
    \label{fig:2g}
  \end{subfigure}

  \caption{\raggedright \textbf{The upper row:} frequency-resolved spectrum of heat dissipation rate of a linear Langevin system with (a) a single delay and (b) two delays. Blue solid lines denote the frequency-resolved spectrum $\tilde{C}_v(\omega) -2T\tilde{R}'_v(\omega)$, green dashed lines show the asymptotic forms: (a) $-2Tb\sin(\omega\tau)/(\gamma^2\omega)$ and (b) $-(2T/\gamma^2\omega)\cdot[A_1\sin(\omega\tau_1) + $ $A_2\sin(\omega\tau_2)]$, and purple crosses denote results of numerical simulations (red error bars show the standard error of the mean). Parameters: (a) $a=-2$, $b=1$, $\tau=1$, $T=0.1$, $\gamma=1$; (b) $A_0 = -3$, $A_1=A_2=1$, $\tau_2 = 2\tau_1 = 2$, $\gamma=1$, $T=0.1$. 
  \textbf{The middle row:} how the single delay time $\tau$ and delay strength $b$ affect the resolved spectrum of heat dissipation rate. We fix $a=-1$, $T=1$, and $\gamma=1$.  
  (c) The spectrum with different delay times $\tau$. We fix $b=0.5$.
  (d) The spectrum with different strengths $b$ of the time-delayed force. Here $\tau=1$. 
  \textbf{The lower row:} how the two delay times $\tau_1, \tau_2$ and corresponding delay strengths $A_1, A_2$ affect the resolved spectrum of heat dissipation rate. We fix $A_0=-1$, $T=1$, and $\gamma=1$.  
  (e) The spectrum with different delay times $\tau_1, \tau_2$. Here $A_1=A_2=0.2$.  
  (f) The spectrum with different delay strengths $A_1,A_2$ of the same sign. Here $\tau_1=1$ and $\tau_2=2$.
  (g) The spectrum with the same delay strengths $A_1,A_2$ of different signs. Here $\tau_1=1$ and $\tau_2=2$.}
  \label{fig:combined}
\end{figure*}

As shown in Fig.~\ref{fig:2a}, throughout the frequency domain, the spectral decomposition of heat dissipation shows a long-lasting oscillation centered at zero with the period $2\pi/\tau$. A longer delay leads to an oscillation of the spectrum with a shorter period [Fig.~\ref{fig:2c}]. The amplitude of the decaying oscillation increases as the magnitude of the strength of the delay term $|b|$ increases, and a negative strength of the delay force ($b<0$) changes the overall sign of the spectrum [Fig.~\ref{fig:2d}]. In particular, in the high-frequency domain ($\omega \gg \vert a\vert/\gamma$  and $\omega \gg |b|/\gamma$), the denominator of $\tilde{C}_v(\omega)-2T\tilde{R}'_v(\omega)$ is dominated by $(\gamma\omega)^2$; hence,
\begin{align}
    & \tilde{R}'_v(\omega)\sim \frac{1}{\gamma} -\frac{b}{ \gamma^2\omega}\sin(\omega\tau)+ O\left(\frac{1}{\omega^2}\right),    \\
    & \tilde{C}_v(\omega)\sim \frac{2T}{\gamma} -\frac{4T}{\gamma^2\omega}b\sin(\omega\tau) + O\left(\frac{1}{\omega^2} \right), \\
    & \tilde{C}_v(\omega)-2T\tilde{R}'_v(\omega)\sim -\frac{2T}{\gamma^2\omega}b\sin(\omega\tau)+
    O\left(\frac{1}{\omega^2}\right).
\end{align}
The asymptotic behavior of these spectra in high frequencies reveals nontrivial information about the time-delayed force: The envelope of the spectral curve decays inversely with frequency $\omega$ and is proportional to the strength $b$ of the time-delayed force. In the vanishing limit of $b$, the entire spectrum tends to zero. This shows an essential nonequilibrium behavior due to the presence of the time-delayed force.

There are two timescales of the dynamics that are important in determining corresponding thermodynamic behavior:
$\tau/\pi$ and $\mathrm{min}\{\gamma/|a|, \gamma/|b|\}$. By sweeping $\omega$ from $0$ to $\infty$, the FRR violation $\tilde{C}_{v}(\omega)-2T\tilde{R}_{v}'(\omega)$ periodically vanishes when $\omega$ is an integer multiple of $\pi/\tau$, which leads to the oscillatory spectrum. In other words, the time-delayed force breaks FRR and contributes to heat dissipation only at timescale $\omega^{-1}$ off-resonant with $\tau/\pi$ divided by an integer. Furthermore, the envelope of the violation spectrum starts to decay around $\omega^{-1} \simeq \mathrm{min}\{\gamma/|a|, \gamma/|b|\}$. This means that the violation of FRR, and hence the contribution to the heat dissipation, is most prominent in timescales longer than the intrinsic relaxation timescale, $\omega^{-1}\geq\mathrm{min}\{\gamma/|a|, \gamma/|b|\}$. For small timescales ($\omega^{-1}\ll\mathrm{min}\{\gamma/|a|, \gamma/|b|\}$), the asymptotic scaling of the spectrum ($\sim 1/\omega$) in the high-frequency region and its oscillation around zero together lead to a negligible contribution to the heat dissipation.

We can preliminarily infer the sign of the heat dissipation and its parameter dependence from the information of the spectrum decomposition in the low-frequency region. Here, we focus on cases with a restoring nondelayed force ($a<0$). Specifically, for a positive delayed force $b>0$, the major contribution to the integral comes from the first valley below zero and the subsequent peak above zero. A shorter delay time results in a longer period of oscillation in the frequency domain, which shifts the first peak towards higher frequency without changing the decay of the envelope [Fig.~\ref{fig:2c}]. Thus, a shorter delay time leads to a lower heat dissipation for $b>0$. In contrast, for negative strength of the delayed force $b<0$, the contribution from the first positive peak of the spectrum dominates, which results in positive heat dissipation [Fig.~\ref{fig:2d} and Fig.~\ref{fig:1}]. Experimentally, these properties allow us to infer the heat dissipation rate of the linear time-delayed systems with relatively small sampling resources.

\section{Linear systems with multidelay times}\label{section:4}

Next, we examine a broader class of time-delayed systems to test the generality of these results. We consider a Langevin equation with multiple delay times:
\begin{equation}\label{multi}
  \left\{
    \begin{alignedat}{2}
      &\gamma \dot{x}(t) = \sum_{i=0}^{N}A_ix(t-\tau_i) + \epsilon f^p(t) + \xi(t) ,
       &(t>t_{\text{init}}), \\ 
      &x(t) = \phi(t), 
      &\hspace{-2.5em}(t\in [t_{\text{init}}-\tau ,t_{\text{init}}]),  \\ 
    \end{alignedat}
  \right.
\end{equation}
where $A_0$ is the strength of a linear force without delay ($\tau_0=0$), $A_i$ for $i=1,\dots,N$ are the strength of a linear time-delayed force term with a constant delay time $\tau_i$ ($\tau_i>0$), and $\tau = \max_i \{\tau_i\}$. Other parameters share the same meaning as in Eq.~\eqref{gen_eq}.
Instead of Eq.~\eqref{convergence_condition1} or~\eqref{convergence_condition2}, we require 
$A_0<-\sum_{i=1}^N |A_i|$ to ensure the existence of a unique stationary solution, which can be shown by the method of Lyapunov functionals~\cite{bailey,Hale1993}. The solution of Eq.~\eqref{multi} is given by 
\begin{align}\label{formal_sol2}
  x(t) &= x_0(t-t_{\text{init}})\phi(t_{\text{init}}) 
  \nonumber \\
  &\quad  + \frac{1}{\gamma}\sum_{i=1}^N A_i\int_{t_{\text{init}}-\tau_i}^{t_{\text{init}}} x_0(t-s-\tau_i)\phi(s)ds \nonumber \\
  & \quad + \frac{1}{\gamma}\int_{t_{\text{init}}}^t x_0(t-s) [ \xi(s) + \epsilon f^p(s) ] ds,
\end{align} 
in which $x_0(t)$ is defined as the solution to Eq.~\eqref{multi} with the conditions in Eq.~\eqref{fundamental}.

Following the same routes as in Sec.~\ref{section:3}, we can obtain the linear response function and the time-correlation function in the steady state as
\begin{align}
    R_v(t) &=\frac{1}{\gamma^2}\sum_{i=0}^{N}A_i x_0(t-\tau_i)+\frac{1}{\gamma}\delta(t),  \label{R_t_multi} \\
    C_v(t) &= \frac{2T}{\gamma^3}\sum_{i,j=0}^{N}A_iA_j K(t+\tau_i-\tau_j)  
    \nonumber \\
    &\quad + \frac{2T}{\gamma^2}\sum_{i=0}^{N}A_i x_0(|t|-\tau_i) +\frac{2T}{\gamma} \delta(t).  \label{C_t_multi}
\end{align}
The Fourier transforms $\tilde{C}_v(\omega)$, $\tilde{R}_v'(\omega)$ and the spectral decomposition of the heat dissipation $\tilde{C}_v(\omega)-2T\tilde{R}'_v(\omega)$ read

\begin{widetext}
\begin{align}
  &\tilde{R}_v'(\omega)= -\frac {1}{\gamma} \frac{\sum_{i,j=0}^{N}A_iA_j\cos(\omega(\tau_i-\tau_j))+\gamma\omega\left[\sum_{i=1}^N A_i\sin(\omega \tau_i)\right]}
  {\left[\sum_{i=0}^N A_i\cos(\omega \tau_i)\right]^2+\left[\gamma\omega+\sum_{i=1}^N A_i\sin(\omega \tau_i)\right]^2} + \frac{1}{\gamma}.  \label{C_v_multi}  \\
  &\tilde{C}_v(\omega) = -\frac{2T}{\gamma} \frac{\sum_{i,j=0}^{N}A_iA_j\cos(\omega(\tau_i-\tau_j))+2\gamma\omega\left[\sum_{i=1}^N A_i\sin(\omega \tau_i)\right]}
  {\left[\sum_{i=0}^N A_i\cos(\omega \tau_i)\right]^2+\left[\gamma\omega+\sum_{i=1}^N A_i\sin(\omega \tau_i)\right]^2}  
    +\frac{2T}{\gamma}.  \label{R_v_multi} \\
  &\tilde{C}_v(\omega)-2T\tilde{R}'_v(\omega) =  -2T \frac{\omega\left[\sum_{i=1}^N A_i\sin(\omega \tau_i)\right]} {\left[\sum_{i=0}^N A_i\cos(\omega \tau_i)\right]^2+\left[\gamma\omega+\sum_{i=1}^N A_i\sin(\omega \tau_i)\right]^2}.  \label{spectrum2}
\end{align}
\end{widetext}
In the high-frequency domain ($\omega \gg |A_0|/\gamma$), we have
\begin{align}
    & \tilde{R}'_v(\omega)\sim \frac{1}{\gamma} -\frac{1}{\gamma^2\omega}\sum_{i=1}^N A_i\sin(\omega \tau_i)+ O\left(\frac{1}{\omega^2}\right),    \\
    & \tilde{C}_v(\omega)\sim \frac{2T}{\gamma} -\frac{4T}{\gamma^2\omega}\sum_{i=1}^N A_i\sin(\omega \tau_i)+ O\left(\frac{1}{\omega^2} \right), \\
    & \tilde{C}_v(\omega)-2T\tilde{R}'_v(\omega)\sim -\frac{2T}{\gamma^2\omega}\sum_{i=1}^N A_i\sin(\omega \tau_i)+
    O\left(\frac{1}{\omega^2}\right).
\end{align}

As an example, we consider systems with two delay times [$N=2$ in Eq.~\eqref{multi}].
Similarly to the case of a single delay, the decomposition spectrum shows an oscillation over the whole frequency domain [Fig.~\ref{fig:2b}]. The amplitude of the oscillation of the spectrum increases as the delay strengths $|A_1|$, $|A_2|$ increase [Fig.~\ref{fig:2f}]; larger delay times $\tau_1$, $\tau_2$ lead to a shorter period of the oscillation [Fig.~\ref{fig:2e}]. In the high-frequency limit, the spectrum becomes a superposition of sinusoidal waves corresponding to different delay times, with a decaying envelope of scaling $1/\omega$.

For systems with two delay times, the contribution to heat dissipation still originates predominantly from the low-frequency region [Fig.~\ref{fig:2b}]. Therefore, we can examine the parameter dependence of the heat dissipation by analyzing the resolved spectrum based on Eq.~\eqref{spectrum2}. For positive strengths of time-delayed forces ($A_1, A_2> 0$), shorter delay times generally yield lower heat dissipation [Fig.~\ref{fig:2e}]. In Fig.~\ref{fig:2f}, positive force strengths ($A_1, A_2 > 0$) can induce negative heat flow, whereas negative strengths ($A_1, A_2<0$) produce positive heat flow; the absolute value of heat dissipation increases with the magnitude of delay strengths of the same sign. Time-delayed forces with opposing signs of the strengths ($A_1>0, A_2<0$ or $A_1<0, A_2>0$) result in higher heat dissipation compared to the both-positive case with the same magnitudes. Notably, flipping the sign (from positive to negative) of the force associated with the shorter delay time generates more dissipation than flipping that of the longer delay time [Fig.~\ref{fig:2g}].

\section{Discussion}
\label{section:conclusion}

In this work, we have studied the spectral decomposition of the heat dissipation rate $\tilde{C}_v(\omega)-2T\tilde{R}'_v(\omega)$ in linear time-delayed Langevin systems using the Harada-Sasa equality. Our main results are summarized as follows: (1) The frequency-resolved spectrum of heat dissipation of linear time-delayed overdamped Langevin systems shows a characteristic oscillation with asymptotic scaling $1/\omega$. 
The magnitude of the oscillation reflects the strength of the delayed force, and its period reflects the delay time. Overall,
such asymptotic behavior reveals nontrivial information about the time-delayed force. (2) The frequency-resolved spectrum in the low-frequency region reflects the sign and the magnitude of the heat dissipation. In experiments, the power spectral density $\tilde{C}_v(\omega)$ can be directly obtained from the recorded trajectories, while the response function $\tilde{R}'_v(\omega)$ is measured by applying a small oscillatory perturbation and analyzing the system’s output (see Appendix~\ref{appendix:numerical} for a brief recipe).
Our study provides a  perspective for identifying and analyzing time-delayed effects in diverse systems, such as active matter, and for effectively quantifying the energetic costs in systems like colloidal particles, electrical circuits, and mechanical oscillators under time-delayed feedback control.

The $1/\omega$ scaling and asymptotic oscillation of frequency-resolved spectrum of heat dissipation are arguably a specific feature of time-delayed systems. This is suggested by a known result for Markovian systems~\cite{time-scale-separation}: For general discrete-state Markovian systems with a general perturbation, the real part of the linear response function, the time correlation function, and the spectrum of heat dissipation approach their respective limits as $\omega\to \infty$ by $1/\omega^2$, as shown by the Taylor expansion of Eq.~(6) in Ref.~\cite{time-scale-separation}. Furthermore, we suspect that the asymptotic oscillations of $\tilde{C}_v(\omega)$ and $\tilde{R}'_v(\omega)$ are not specific to linear systems but may also arise in some nonlinear time-delayed systems (e.g., see~\cite{kopp}). A systematic verification of this behavior in a broader class of nonlinear time-delayed stochastic systems is an interesting direction for future theoretical investigation. 

Additionally, time-delayed systems are often analyzed using enlarged models that explicitly incorporate additional degrees of freedom (for instance, through Markovian embedding using the linear chain trick~\cite{delay_textbook,sarah2,rosinberg3}, or by directly modeling the feedback mechanism as an explicit degree of freedom~\cite{rosinberg1}). Understanding our results using such enlarged models is also an interesting direction for gaining further physical insights into these characteristics.

\begin{acknowledgments}
The authors thank Naomichi Hatano for careful reading of the manuscript. N.~O.~thanks Sosuke Ito for discussions. X.W.~acknowledges financial support from the Department of Physics, Graduate School of Science, the University of Tokyo. N.O.~was supported by JSPS KAKENHI Grant No.~23KJ0732. R.B.~was supported by JSPS KAKENHI Grant No. 25KJ0766.
\end{acknowledgments}

\appendix

\section{Numerical methods}\label{appendix:numerical}
We briefly sketch the method of numerical calculation. The procedure is in principle applicable in experimental setups as well. 

For a single trajectory $x(t)$ at resolution $\delta t$, we assume the steady state after a sufficiently long time $t=t_0$ and take discrete Fourier transform (DFT) of the data in $[t_0, t_0 + \mathcal{T}] \ (\mathcal{T}=N_{\mathrm{sp}}\delta t)$, as suggested in the Supplemental Material of Ref.~\cite{time-scale-separation}:
\begin{equation}
\tilde{x}(\omega_k) = \frac{\sqrt{\delta t}}{\sqrt{N_{\mathrm{sp}}}} \sum_{j=1}^{N_{\mathrm{sp}}}x(t_0+j\delta t)\exp[-i\omega_k (t_0+j\delta t) ], 
\end{equation}
where $\omega_k = 2\pi k /N_{\mathrm{sp}}\delta t  \ \  (k = 0, \cdots , N_{\mathrm{\mathrm{sp}}}-1)$. The power spectral density of velocity $\tilde{C}_v(\omega)$ is obtained by 
\begin{equation}
    \tilde{C}_v(\omega)=\langle|\tilde{v}(\omega)|^2\rangle=\omega^2\langle|\tilde{x}(\omega)|^2\rangle,
\end{equation}
where $\langle\cdots\rangle$ represents the ensemble average over a sufficiently large number of trajectories. 
For the response function $\tilde{R}_v'(\omega_0)$ for a specific $\omega_0$, first, we plug  $f^p(t)=e^{i\omega_0 t}$  into Eq.~\eqref{response}:
\begin{align}\label{response_complex}
\langle \dot{x}(t)\rangle &= v_s + \epsilon \int_{-\infty}^{t}R_v(t-s)e^{i\omega_0 s}ds +O(\epsilon^2) \notag \\
&= v_s+ \epsilon \tilde{R}_v(\omega_0)e^{i\omega_0 t}+O(\epsilon^2).   
\end{align}
It is more practical to consider the response to a real perturbation $f^p(t)=\cos(\omega_0t) = \mathrm{Re} [e^{i\omega_0 t}]$, which is given by taking the real part of Eq.~\eqref{response_complex} ($v_s=0$ in the steady state and ignoring $O(\epsilon^2)$):
\begin{equation}
\langle \dot{x}(t)\rangle = \epsilon\{ \text{Re}[R_v(\omega_0)]\cos(\omega_0 t)-\text{Im}[R_v(\omega_0)]\sin(\omega_0 t)\},
\end{equation}
which is equivalent to 
\begin{align}
\langle {x}(t)\rangle &= \frac{\epsilon}{\omega_0}\{ \text{Re}[R_v(\omega_0)]\sin(\omega_0 t)+\text{Im}[R_v(\omega_0)]\cos(\omega_0 t)\}   \nonumber \\
& \ \ + \mathrm{constant}.
\end{align}
The real part of $\tilde{R}_v(\omega_0)$ is extracted by taking the inner product of averaged trajectories $\langle x(t)\rangle$ in $[t_0,t_0+\mathcal{T}]$ with $\sin(\omega_0t)$:
\begin{equation}
    \text{Re}[R_v(\omega_0)] \approx  \frac{\omega_0}{\epsilon}\cdot  \frac{2}{N_\mathrm{sp}}\sum_{j=1}^{N_\mathrm{sp}}\langle x(t_0+j\delta t)\rangle\sin[\omega_0(t_0+j\delta t)].
\end{equation}

In numerical simulations, we compute $\omega^2|\tilde{x}(\omega)|^2$ for each trajectory without any perturbation force and then take the average $\tilde{C}_v(\omega)$ and the standard deviation $\sigma_C$ over an ensemble of $N$ trajectories. Similarly, for a specified frequency $\omega_0$, we add a perturbation $f^p(t)=\cos(\omega_0t)$ and compute the expression $2\omega_0 (\epsilon N_{\mathrm{sp}})^{-1}
\sum_{j=1}^{N_{\mathrm{sp}}} x(t_0 +j\delta t) \sin[\omega_0 (t_0 +j\delta t)]$ for each trajectory. 
Then we take the average as $\tilde{R}'_v(\omega_0)$ and the standard deviation $\sigma_{R'}$ over $N$ trajectories.  
Since $\sigma_C$ and $\sigma_{R'}$ are computed from different dynamical processes and noise realizations, the error bars for the spectrum $\tilde{C}_v(\omega)-2T\tilde{R}'_v(\omega)$ are given by $\sqrt{(\sigma_C^2+4T^2\sigma_{R'}^2)/N}$.

We use the programming language \textit{Julia} and the package \textit{DifferentialEquations.jl}~\cite{julia_dde} and \textit{StochasticDelayDiffEq.jl}~\cite{julia_sdde} for numerical simulations using the Euler-Maruyama method~\cite{EM}. For $\tilde{C}_v(\omega)$ or $\tilde{R}'_v(\omega_0)$ (with a selected $\omega_0$), we set the steady-state time $t_0 = 200$ and simulate for a duration of $\mathcal{T} = 500$
with a resolution of $\delta t = 0.001$ over $N=50,000$ trajectories in both single  and two-delay cases.

\section{Proof of the Harada-Sasa equality for systems with time-delayed forces}
\label{appendix:Harada-Sasa}

We prove the Harada-Sasa equality in Eq.~\eqref{hs} based on Refs.~\cite{harada-sasa2,ito}. For convenience, we write the external force at time $t$ as
\begin{equation}
    \mathcal{F}_\epsilon (t) \equiv  F(x(t),x(t-\tau_1),\cdots,x(t-\tau_n ))+ \epsilon f^p(t),
\end{equation}
which is a stochastic quantity.

We first consider overdamped systems ($m=0$) and use the path-integral formalism~\cite{path-integral,Shiraishi}. We consider discrete time points $t_i\equiv t_\mathrm{init}+i\Delta t$ for $i=0,1,\dots$ with a small interval $\Delta t$. We take the following discrete time evolution, which converges to the original time evolution in Eq.~\eqref{gen_eq} in the $\Delta t\to0$ limit:
\begin{equation}
    \label{langevin-discrete}
    \gamma \Delta x^{i} = \mathcal F^i_\epsilon \Delta t + \sqrt{2\gamma T} \Delta W^i,
\end{equation}
where $\Delta x^i \equiv x(t_{i+1}) - x(t_i)$, $\mathcal{F}^i_\epsilon \equiv \mathcal F_\epsilon(t_i)$, and $\Delta W^i$ is the Wiener increment obeying the Gaussian distribution with mean $\langle \Delta W^i\rangle =0$ and correlation $\langle \Delta W^i\Delta W^j \rangle=\delta_{ij}\Delta t$ ($\sqrt{2\gamma T}\Delta W^i/\Delta t\simeq\xi(t_i)$). Given an initial history $\{\phi(s)\,\vert\,t_\mathrm{init}-\tau\leq s\leq t_\mathrm{init}\}$, the external force $\mathcal{F}^i_\epsilon$ is regarded as a function of $\{\Delta x^0,\dots,\Delta x^{i-1}\}$.

We consider a path from time $t_\mathrm{init} \equiv t_0 $ to $t_N$. Given an initial condition $\{ \phi(s) \,\vert\, t_\mathrm{init}-\tau \leq s \leq t_\mathrm{init}\} $, a realization $\{\Delta x^0,\dots,\Delta x^{N-1}\}$ is completely determined by the realization of the noise $\{\Delta W^0,\dots,\Delta W^{N-1}\}$. The noise obeys a multivariate Gaussian distribution, and thus the probability density of a path $\{\Delta x^0,\dots,\Delta x^{N-1}\}$ is given by
\begin{align}
    &P\bigl[\{\Delta x^0,\dots,\Delta x^{N-1}\}\,\vert\,\{\phi(\cdot)\}\bigr] 
    \nonumber \\
    &\quad = \frac{\mathcal{J}}{\mathcal{N}} \exp\left\{-\frac{1}{4\gamma T\Delta t}\sum_{i=0}^{N-1}[\gamma \Delta x^{i}-\mathcal F^i_\epsilon\Delta t]^2\right\},
\end{align}
where 
\begin{equation}
    \mathcal{J} = \left\vert \det\left[ \frac{\partial (\Delta W^0,\dots,\Delta W^{N-1})}{\partial (\Delta x^0,\dots,\Delta x^{N-1})}\right] \right\vert
\end{equation}
is the Jacobian and $\mathcal{N} = (\sqrt{2\pi\Delta t})^N$ is the normalization constant. The Jacobian matrix consists of the elements $\partial (\Delta W^i)/\partial (\Delta x^j) $. Using Eq.~\eqref{langevin-discrete}, we can find $\partial (\Delta W^i)/\partial (\Delta  x^j) =0$ for $j > i$, and thus the Jacobian matrix is a lower triangular matrix. Thus, the determinant is the product of the diagonal elements, $\prod_{i=0}^{N-1} \partial (\Delta W^{i})/\partial (\Delta x^i) = (\sqrt{\gamma /2T})^N$. Notably, this determinant does not depend on the perturbation force $\epsilon f^p(t)$.

Taking the continuous limit ($\Delta t\rightarrow0$), the probability density of a path  $\{x(s)\,\vert\, t_\mathrm{init} < s \leq t \}$ is given by the path integral
\begin{align}
    &P\bigl[ \{x(\cdot )\} \,\vert\, \{\phi(\cdot )\}\bigr] 
    = \frac{\mathcal{J}}{\mathcal{N}} \exp[-\mathcal{S}_\epsilon(\{x\})],
    \\
    & \mathcal{S}_\epsilon(\{x\})\equiv \int_{t_\mathrm{init}}^t \frac{1}{4\gamma T}[\gamma \dot x(s) - \mathcal F_\epsilon(s)]^2ds,
\end{align}
with $\mathcal{N}$ and $\mathcal{J}$ independent of $\epsilon f^p(t)$.

Now we use this path integral to prove the Harada-Sasa equality. For $s'\neq s$, we can rewrite the response function as
\begin{align}
    \label{R_derive}
    &R_v(s'-s)
    \equiv \frac{\delta \left\langle \dot{x} (s')\right\rangle_\epsilon }{\delta \epsilon f^p(s)} \Bigg|_{\epsilon=0}  \nonumber \\
    &  = \frac{\delta}{\delta \epsilon f^p(s)}\left[ \frac{\mathcal J}{\mathcal N} \int \mathcal D\{x\}\, e^{-\mathcal S_\epsilon(\{x\})} \dot{x}(s')\right] \Bigg|_{\epsilon=0}
    \nonumber \\
    &  = - \frac{\mathcal J}{\mathcal N} \int \mathcal D\{x\}\,  e^{-\mathcal S_0(\{x\})} \dot{x}(s') \frac{\delta \mathcal{S}_\epsilon(\{x\})}{\delta \epsilon f^p(s)}\Biggl|_{\epsilon=0}
    \nonumber \\
    &  = \frac{\mathcal J}{\mathcal N} \int \mathcal D\{x\}\, e^{-\mathcal{S}_0(\{x\})} \dot{x}(s') \frac{1}{2\gamma T} [\gamma \dot{x}(s) - \mathcal{F}_0(s) ]  
    \nonumber \\
    & = \frac{1}{2\gamma T} \langle \dot{x}(s') [\gamma \dot{x}(s) - \mathcal{F}_0(s) ] \rangle_0.
\end{align}
Combining this with the definition of the correlation function in Eq.~\eqref{corr_fun}, we obtain
\begin{align}
    \label{C-R_derive}
    & C_v(s'-s) - 2T R_v(s'-s) 
    \nonumber \\
    &= \langle [\dot{x}(s') - v_s][\dot{x}(s) -v_s] \rangle_0
    - \frac{1}{\gamma} \langle \dot x (s')[\gamma \dot{x}(s) - \mathcal{F}_0(s) ]\rangle_0  
    \nonumber \\
    &= \frac{1}{\gamma} \langle \dot{x}(s') \mathcal{F}_0(s) \rangle_0 - v_s^2.
\end{align}
So far, we have performed calculations for $s'\neq s$. 
We use this relation to rewrite the heat dissipation rate as
\begin{align}
  \label{J_derive}
  & \left\langle J \right\rangle_0 
  = \left\langle \mathcal{F}_0(s) \circ \dot{x}(s)\right\rangle_0  
  \nonumber \\
  &= \lim_{t\to 0^+} \frac{1}{2}
  \langle \mathcal{F}_0(s+t)\dot{x}(s) +\mathcal{F}_0(s-t) \dot{x}(s) \rangle_0 
  \nonumber \\
  &= \gamma v_s^2+\lim_{t\to 0^+} \frac{\gamma}{2}\bigl\{[C_v(t)-2TR_v(t)] + [C_v(-t)-2TR_v(-t)] \bigr\} . 
\end{align}
On the other hand, the singularity of $C_v(0)$ is given by $\gamma^{-2} \langle \xi(s)\xi(s) \rangle = (2 T/\gamma ) \delta(0)$ and the singularity of $R_v(0)$ is $\gamma^{-1} \delta [\epsilon f^p (s)]/\delta [\epsilon f^p (s)] = \gamma^{-1} \delta (0)$. Thus, the combination $C_v(0) - 2T R_v(0)$ has no singularity. Therefore, we can apply the equality $\int_{-\infty}^{\infty}\tilde{f}'(\omega)\frac{d\omega}{2\pi}=\frac{1}{2}\lim_{t\rightarrow 0^+}\left[ f(t) + f(-t) \right]$ to $f(t) = C_v(t) - 2T R_v(t)$. This leads to the Harada-Sasa equality in Eq.~\eqref{hs}.

For underdamped systems ($m\neq 0$), we similarly consider the path integral by taking the velocity $v(s) \equiv \dot{x}(s)$ as the variable. Given an initial condition $\{ \phi(s) \,\vert\, t_\mathrm{init}-\tau \leq s \leq t_\mathrm{init}\} $ and an initial velocity $v_0$, the path of the velocity $\{v(s)\,\vert\,t_\mathrm{init}<s\leq t\}$ completely determines the path of the position $\{x(s)\,\vert\,t_\mathrm{init}<s\leq t\}$ and hence the time-delayed force. By a similar procedure as above, we find the path integral expression of the probability density,
\begin{align}
    &P\bigl[ \{v(\cdot )\} \,\vert\, \{\phi(\cdot )\},v_0\bigr] 
    = \frac{\mathcal{J}}{\mathcal{N}} \exp[-\mathcal{S}_\epsilon(\{v\})],
    \\
    & \mathcal{S}_\epsilon(\{v\})\equiv \int_{t_\mathrm{init}}^t \frac{1}{4\gamma T}[m\dot v(s)+\gamma v(s) - \mathcal F_\epsilon(s)]^2ds,
\end{align}
with the Jacobian $\mathcal{J}$ and the normalization $\mathcal{N}$ independent of $\epsilon f^p(\cdot)$. 

We can prove the Harada-Sasa equality for underdamped systems similarly to Eqs.~\eqref{R_derive}--\eqref{J_derive}. The response function is calculated from the path integral as
\begin{equation}
    R_v(s'-s) = \frac{1}{2\gamma T} \bigl\langle v (s') [m\dot{v}(s) + \gamma v(s) - \mathcal{F}_0(s) ] \bigr\rangle_0.
\end{equation}
Combining this with the definition of the correlation function, we obtain
\begin{align}
    &C_v(s'-s) - 2T R_v (s'-s) 
    \nonumber \\
    &\quad = \frac{1}{\gamma} \bigl\langle v(s')[-m\dot{v}(s) + \mathcal F_0(s)] \bigr\rangle_0 - v_s^2.
\end{align}
The heat dissipation rate is thus given by
\begin{align}
  &\left\langle J \right\rangle_0 
  = \bigl\langle [-m\dot v(s)+\mathcal{F}_0(s)] \circ v(s)\bigr\rangle_0  
  \nonumber \\
  &=\lim_{t\to 0^+} \frac{1}{2}
  \bigl\langle [- m\dot v(s) + \mathcal{F}_0(s)][v(s+t)+v(s-t)] \bigr\rangle_0 
  \nonumber \\
  &= \gamma v_s^2+\lim_{t\to 0^+} \frac{\gamma }{2}\bigl\{[C_v(t)-2TR_v(t)] + [C_v(-t)-2TR_v(-t)] \bigr\} .
\end{align}
Since there is no singularity of delta functions for underdamped systems, we can readily use the equality $\int_{-\infty}^{\infty}\tilde{f}'(\omega)\frac{d\omega}{2\pi}=\frac{1}{2}\lim_{t\rightarrow 0^+}\left[ f(t) + f(-t) \right]$ to reproduce the Harada-Sasa equality.

\section{Derivations of \texorpdfstring{$C_v(t)$, $R_v(t)$, and $\langle J\rangle_0$}{Cv(t), Rv(t), and J0}}\label{appendix:CR_t}

We directly calculate the linear response function and the time correlation function of the linear time-delayed overdamped Langevin systems with a single time-delay term [Eq.~\eqref{par_eq}]. From the formal solution Eq.~\eqref{formal_sol}, the velocity is given by
\begin{align}
\label{x_dot}
  &\dot{x}(t) = (\text{historical terms})  \nonumber \\ 
   &\quad + \frac{1}{\gamma^2}\int_{t_{\text{init}}}^t 
  [ax_0(t-s)+bx_0(t-\tau-s) ][ \xi(s) + \epsilon f^p(s) ]ds  \nonumber \\
  &\quad + \frac{1}{\gamma} [\xi(t) + \epsilon f^p(t) ].
\end{align}
Taking the limit $t_{\text{init}}\rightarrow -\infty$, 
the historical terms vanish. By taking the average $\langle \cdots  \rangle_0$, we find that the steady-state velocity vanishes: $v_s = \langle \dot{x}(t)\rangle _0=0$.

By taking the functional derivative of Eq.~\eqref{x_dot}, we obtain the linear response function at steady state,
\begin{align}\label{R_explicit}
  R_v(t) &\equiv \frac{\delta \left\langle \dot{x} (s)\right\rangle_\epsilon }{\delta \epsilon f^p(s-t)}\bigg|_{\epsilon=0}  \nonumber \\
  &= \frac{1}{\gamma^2}[ ax_0(t)+bx_0(t-\tau)]+ \frac{1}{\gamma}\delta(t)
\end{align}
for all $t$. In particular, we have $R_v(t)=0$ for $t<0$.

Next, using Eq.~\eqref{x_dot} and the steady-state velocity $v_s = 0$, the time correlation function in the steady state is expanded as
\begin{align}
\label{C_v_mid0}
  & C_v(t) \equiv \bigl\langle [\dot{x}(s)-v_s] [\dot{x}(s+t) - v_s] \bigr\rangle _0
  \nonumber \\
  &=  \left\langle \dot{x}(s) \dot{x}(s+t) \right\rangle_0
  \nonumber \\
  &= \Biggl\langle \left( \frac{1}{\gamma^2}\! \int_{-\infty}^s [ax_0(s-s')+bx_0(s-\tau-s') ]\xi(s')ds' + \frac{\xi(s) }{\gamma} 
  \right)    \nonumber \\
  &\,\,\,\, \times   \Biggl(\frac{1}{\gamma^2} \! \int_{-\infty}^{s+t} [ ax_0(s+t-s') 
   +bx_0(s+t-\tau-s') ] \xi(s')ds'\nonumber \\ & 
   \quad \quad \quad +\frac{\xi(s+t) }{\gamma} \Biggr)  \Biggr\rangle \nonumber \\
  &=\frac{2T}{\gamma^3}\! \int_{-\infty}^{s} [ ax_0(s-s')+bx_0(s-\tau-s') ] \nonumber \\
  &\quad \quad \quad \quad \times [ ax_0(s+t-s')+bx_0(s+t-\tau-s') ] ds'   \nonumber \\ 
  &\quad + \frac{2T}{\gamma^2} [ ax_0(\vert t\vert)+bx_0(\vert t \vert-\tau)] +\frac{2T}{\gamma}\delta(t),
\end{align} 
where we neglect additional finite terms that are only present at $t=0$ because they will not contribute to the Fourier transform of $C_v(t)$. Noting that $x_0(t)=0$ for $t<0$, we can replace the integral $\int_{-\infty}^s ds'$ in the last expression with $\int_{-\infty}^{\infty}ds'$. Defining
\begin{equation}
    \label{K}
  K(t) 
  \equiv \int_{0}^{\infty}x_0(s)x_0(s+t)ds
  = \int_{-\infty}^{\infty}x_0(s)x_0(s+t)ds,
\end{equation}
which satisfies $K(t)=K(-t)$, we can rewrite Eq.~\eqref{C_v_mid0} as
\begin{align}\label{C_explicit}
  C_v(t)&=\frac{2T}{\gamma^3}\left\{ ( a^2 + b^2 )  K(t) + ab [ K(t+\tau) + K(t-\tau) ] \right\}  \nonumber \\
  & \quad + \frac{2T}{\gamma^2} [ ax_0(\vert t\vert)+bx_0(\vert t \vert-\tau)] +\frac{2T}{\gamma}\delta(t)
\end{align}
for all $t$. By similar calculations, we can obtain the expressions of $C_v(t)$ and $R_v(t)$ for the systems with multiple delay times in Eqs.~\eqref{R_t_multi} and \eqref{C_t_multi}.

The steady-state heat dissipation rate for the system with a single delay time $\tau>0$ [Eq.~\eqref{par_eq}] can be calculated as 
\begin{align}
\label{J_0}
  \left\langle J \right\rangle_0 &\equiv  \left\langle[\gamma \dot{x}(t)-\xi(t)]\circ \dot{x}(t)\right\rangle
  \nonumber \\
  &=  \left\langle[ax(t)+bx(t-\tau)]\circ \dot{x}(t)\right\rangle
  \nonumber \\
     &=  \left\langle[ax(t)+bx(t-\tau)]\circ \frac{1}{\gamma}\{ [ax(t)+bx(t-\tau)]+\xi(t)  \}  \right\rangle
  \nonumber \\
  &= \Biggl\langle \left( \frac{1}{\gamma}\! \int_{-\infty}^s [ax_0(s-s')+bx_0(s-\tau-s') ]\xi(s')ds' \right)    \nonumber \\
  &\,\,\,\, \circ   \Biggl(\frac{1}{\gamma^2} \! \int_{-\infty}^{s} [ ax_0(s-s') 
   +bx_0(s-\tau-s') ] \xi(s')ds'\nonumber \\ & 
   \quad \quad \quad +\frac{\xi(s) }{\gamma} \Biggr)  \Biggr\rangle \nonumber \\
  &=\frac{T}{\gamma} a+ \frac{2T}{\gamma^2}[ ( a^2 + b^2 )  K(0) + 2ab K(\tau) ],
\end{align} 
where we used $\bigl\langle \bigl[\int^s_{-\infty} f(s') \xi(s')ds' \bigr]\circ \xi (s) \bigr\rangle = \frac{1}{2} (2\gamma T) f(s) $ due to the Stratonovich product.
We also used $\langle \xi(t-s)\xi(t)\rangle=0$ for $s\geq\tau$, and accordingly $\langle x(t-\tau)\xi(t)\rangle=0$. 
The quantities $K(0)$ and $K(\tau)$ have the following analytical expressions under the conditions Eq.~\eqref{convergence_condition1} or Eq.~\eqref{convergence_condition2}~\cite{kuchler1992}:
\begin{align}\label{K_tau}
  K(0)&= \left\{
    \begin{alignedat}{2}
      & \gamma\cdot \dfrac{b\sinh(l\tau/\gamma)-l}{2l(a+b\cosh(l\tau/\gamma))},  \quad \quad  &&(|b|<-a)\\    
      &\dfrac{b\tau-\gamma}{4b}, &&(b=a)  \\ 
      &\gamma\cdot\dfrac{b\sin(l\tau/\gamma)-l}{2l(a+b\cos(l\tau/\gamma))},  &&(b<-|a|)\\ 
    \end{alignedat}
  \right.
  \\
  \label{K_0}
  K(\tau)&= \left\{
    \begin{alignedat}{2}
      & K(0)\cdot \cosh(l\tau/\gamma)-\frac{\gamma\sinh(l\tau/\gamma)}{2l},  \quad  &&(|b|<-a)\\    
      &K(0)-\frac{\tau}{2}, &&(b=a)  \\ 
      &K(0)\cdot \cos(l\tau/\gamma)-\frac{\gamma\sin(l\tau/\gamma)}{2l},  &&(b<-|a|)\\ 
    \end{alignedat}
  \right.
\end{align}
where we define $l=\sqrt{|a^2-b^2|}$ and note that $a$ and $b$ in Ref.~\cite{kuchler1992} correspond to $a/\gamma$ and $b/\gamma$ in this paper. In the $\tau\to0$ limit, $K(0)$ approaches $-\gamma/[2(a+b)]$  and $\langle J\rangle_0$ approaches $-bT/\gamma$.

Note that when $\tau=0$, $\langle bx(t-\tau)\circ\xi(t)/\gamma\rangle$ in the third line of Eq.~\eqref{J_0} will contribute an additional term $bT/\gamma$, leading to $\langle J\rangle_0= -bT/\gamma  + bT/\gamma =0$. This abrupt emergence of the additional term is due to the mathematical assumption of white noise $\langle \xi(t-\tau) \xi(t)\rangle \propto \delta(\tau)$. Physically, the white-noise assumption becomes invalid when $\tau$ is as short as the timescale of the environmental noise correlation, and this additional term should gradually emerge when $\tau$ approaches zero.

\section{Derivations of \texorpdfstring{$\tilde{C}_v(\omega)$ and $\tilde{R}_v(\omega)$}{Cv(omega) and Rv(omega)}}
\label{appendix:CR_omega}

We calculate the Fourier transform of linear response function $R_v(t)$ and the time correlation function $C_v(t)$ for systems with a single delay time. The Fourier transform of $R_v(t)$ is 
\begin{align}
\label{R_v_omega_mid0}
  \tilde{R}_v(\omega) &= \int_{-\infty}^{\infty} \left\{ \frac{1}{\gamma^2} [ax_0(t) + bx_0(t-\tau)]+ \frac {1}{\gamma} \delta(t) \right\} e^{i\omega t}dt
  \nonumber \\
  &= \frac{1}{\gamma^2}(a+be^{i\omega\tau})\tilde{x}_0(\omega)+\frac{1}{\gamma}.
\end{align}
Here, $\tilde{x}_0(\omega)$ is the Fourier transform of the fundamental solution, which can be calculated as follows.
Recalling that the fundamental solution $x_0(t)$ is the solution of Eq.~\eqref{par_eq} with the conditions in Eq.~\eqref{fundamental}, we have
\begin{equation}\label{fun_eq_single}
     \dot{x}_0(t) = \frac{1}{\gamma}\left[ ax_0(t)+bx_0(t-\tau)\right] + \delta(t),
\end{equation}
where the delta function arises from the discontinuity of $x_0(t)$ at $t=0$ from $0$ to $1$ due to the choice of the historic trajectory $\phi(t)$ in Eq.~\eqref{fundamental}. Using this relation, we obtain
\begin{align}\label{x_0_der}
    &\frac{\partial }{\partial t}[x_0(t)e^{i\omega t}]   \nonumber \\
    &\quad= \frac{1}{\gamma}\left[ ax_0(t)+bx_0(t-\tau)\right]e^{i\omega t} + i\omega x_0(t)e^{i\omega t}+\delta(t)e^{i\omega t}.
\end{align}
Integrating Eq.~\eqref{x_0_der} from $t=-\infty$ to $t=\infty$ and using the definition of the Fourier transform gives 
\begin{align}
    &\big[x_0(t)e^{i\omega t}\big]\bigg|_{t=-\infty}^{t=\infty} \nonumber \\
    &\quad= \frac{1}{\gamma}\left[ a\tilde{x}_0(\omega)+b\tilde{x}_0(\omega)e^{i\omega \tau}\right]  +i\omega \tilde{x}_0(\omega) + 1.
\end{align}
The left-hand side vanishes because $x_0(t)=0$ for $t<0$ and $\lim_{t\to\infty} x_0(t)=0$. Rearranging gives
\begin{equation}\label{x_0_omega}
    \tilde{x}_0(\omega) = -\frac{\gamma}{[a + b \cos(\omega\tau)]+ i[\gamma \omega+b\sin(\omega\tau)]}.
\end{equation}
Inserting Eq.~\eqref{x_0_omega} into Eq.~\eqref{R_v_omega_mid0} gives
\begin{align}\label{R_v_omega_result}
  &\tilde{R}'_v(\omega) = \text{Re}\left[ \tilde{R}_v(\omega)  \right] 
  \nonumber \\
  & =\frac{1}{\gamma^2}\text{Re}\left[(a+be^{i\omega\tau})  \tilde{x}_0(\omega) \right] +\frac{1}{\gamma} \nonumber \\
  &= -\frac{1}{\gamma}\frac{[a+b\cos(\omega\tau)]^2+b\sin(\omega\tau)[\gamma\omega+b\sin(\omega\tau)]}{[a+b\cos(\omega\tau)]^2+[\gamma\omega+b\sin(\omega\tau)]^2} +\frac{1}{\gamma}
  \nonumber \\
  &= -\frac {1}{\gamma} \frac{a^2+b^2 + 2ab\cos(\omega\tau) +b\gamma\omega \sin(\omega\tau) }{[a+b\cos(\omega\tau)]^2+[\gamma\omega+b\sin(\omega\tau)]^2} + \frac{1}{\gamma}.
\end{align}

Next, the Fourier transform of the time correlation function in Eq.~\eqref{C_explicit} reads
\begin{align}\label{C_v_omega_mid0}
  & \tilde{C}_v(\omega) \nonumber \\
  & = \frac{2T}{\gamma^3}\! \int_{-\infty}^{\infty} \Bigl\{  (a^2+b^2) K(t) + ab\left[ K(t+\tau) + K(t-\tau) \right]  \Bigr\}
   e^{i\omega t}dt  \nonumber \\
  &\quad + \frac{4T}{\gamma^2} \mathop{\mathrm{Re}}\left[ \int_0^{\infty}\big[ax_0(t)+bx_0(t-\tau)\big]e^{i\omega t}dt\right]  + \frac{2T}{\gamma} \nonumber \\
  &= \frac{2T}{\gamma^3} \bigl[ (a^2+b^2) + ab(e^{-i\omega\tau} + e^{i\omega\tau}) \bigr] \tilde{K}(\omega) \nonumber \\
  &\quad +\frac{4T}{\gamma ^2}\mathrm{Re}\bigl[(a+be^{i\omega \tau })\tilde{x}_0(\omega)\bigr]  + \frac{2T}{\gamma}.
\end{align}
We have already calculated $\mathrm{Re}[(a+be^{i\omega \tau })\tilde{x}_0(\omega)]$ in Eq.~\eqref{R_v_omega_result}. We need to calculate the Fourier transform of $K(t)$. We rewrite the definition of $K(t)$ in Eq.~\eqref{K} using the inverse Fourier transform $x_0(t) = \int_{-\infty}^\infty \tilde{x}_0(\omega) e^{-i\omega t} \frac{d\omega}{2\pi} $ as
\begin{align}
    &K(t) = \int_{-\infty}^{\infty} x_0(s)x_0(s+t)ds
    \nonumber \\[-0.2em]
    &= \int_{-\infty}^\infty ds \int_{-\infty}^\infty \frac{d\omega}{2\pi} \int_{-\infty}^\infty \frac{d\omega'}{2\pi} 
    \tilde{x}_0(\omega')e^{-i\omega' s} \tilde{x}_0(\omega)e^{-i\omega (s+t)} 
    \nonumber \\
    &= \int_{-\infty}^\infty \frac{d\omega}{2\pi} \int_{-\infty}^\infty \frac{d\omega'}{2\pi} \int_{-\infty}^\infty ds\, 
    \tilde{x}_0(\omega') \tilde{x}_0(\omega) e^{-i(\omega' + \omega) s}e^{-i\omega t} 
    \nonumber \\
    &=  \int_{-\infty}^\infty \frac{d\omega}{2\pi} \int_{-\infty}^\infty d\omega' 
    \tilde{x}_0(\omega') \tilde{x}_0(\omega) \delta (\omega' + \omega)  e^{-i\omega t} 
    \nonumber \\
    &= \int_{-\infty}^\infty \frac{d\omega}{2\pi} \tilde{x}_0(-\omega) \tilde{x}_0(\omega) e^{-i\omega t}.
\end{align}
\vspace{1.0em}
Comparing this expression with $K(t) = \int_{-\infty}^\infty  \tilde{K}(\omega) e^{-i\omega t} \frac{d\omega}{2\pi}$, we obtain the Fourier transform of $K(t)$,
\begin{align}
    \tilde{K}(\omega) 
    &= \tilde{x}_0(-\omega) \tilde{x}_0(\omega)
    \nonumber \\
    &=\frac{\gamma^2} { [a+b\cos(\omega\tau)]^2 + [\gamma\omega+b\sin(\omega\tau)]^2 }.
\end{align}
Inserting into Eq.~\eqref{C_v_omega_mid0} gives
\begin{align}
    \label{C_v_omega_result}
    &\tilde{C}_v(\omega) 
    \nonumber \\
    &= \frac{2T}{\gamma} \frac{a^2 + b^2 + 2ab\cos (\omega \tau)} { [a+b\cos(\omega\tau)]^2 + [\gamma\omega+b\sin(\omega\tau)]^2 } 
    \nonumber \\
    &\quad - \frac{4T}{\gamma}\frac{a^2+b^2 + 2ab\cos(\omega\tau) +b\gamma\omega \sin(\omega\tau)} {[a+b\cos(\omega\tau)]^2+[\gamma\omega+b\sin(\omega\tau)]^2} 
    +\frac{2T}{\gamma}
    \nonumber \\
    & = -\frac{2T}{\gamma} \frac{a^2 + b^2 + 2ab \cos(\omega\tau) + 2b\gamma\omega \sin(\omega \tau )} {[a+b\cos(\omega\tau)]^2+[\gamma\omega+b\sin(\omega\tau)]^2}  
    +\frac{2T}{\gamma}.
\end{align}

For the dynamics with multiple delay times, 
we assume the condition $A_0<-\sum_{i=1}^N |A_i|$, so that $\lim_{t\rightarrow\infty}x_0(t)=0$. Following the same approach used for the single-delay case, the Fourier transform of the fundamental solution $x_0(t)$ of Eq.~\eqref{multi} can be expressed as
\begin{equation}\label{x_o_multi}
     \tilde{x}_0(\omega) = -\dfrac{\gamma}{i\gamma\omega + \sum_{i=0}^{N}A_ie^{i\omega \tau_i}}.
\end{equation}
Then,
\begin{align}
    \tilde{K}(\omega) &= \tilde{x}_0(\omega)\tilde{x}_0(-\omega)  \nonumber \\
    & = \dfrac{\gamma^2}{\bigl[\sum_{i=0}^{N}A_i\cos(\omega\tau_i)\bigr]^2+\bigl[\gamma\omega+\sum_{i=1}^{N}A_i\sin(\omega\tau_i)\bigr]^2},
\end{align}
where we use $\sin(\omega \tau_0)=0$.
By a similar calculation to the case of a single delay time, we conclude Eqs.~\eqref{C_v_multi} and \eqref{R_v_multi}.

\vspace{4\baselineskip}

%

\end{document}